\documentclass{article}%
\usepackage{amssymb}
\usepackage{amsfonts}
\usepackage{amsmath}%
\usepackage{graphicx}

\begin{document}

\title{Rao-Blackwellized Particle Smoothing as Message Passing}

\maketitle

\begin{abstract}
In this manuscript the fixed-lag smoothing problem for conditionally linear
Gaussian state-space models is investigated from a factor graph perspective.
More specifically, after formulating Bayesian smoothing for an arbitrary
state-space model as forward-backward message passing over a factor graph, we
focus on the above mentioned class of models and derive a novel
Rao-Blackwellized particle smoother for it. Then, we show how our technique
can be modified to estimate a point mass approximation of the so called joint
smoothing distribution. Finally, the estimation accuracy and the computational
requirements of our smoothing algorithms are analysed for a specific
state-space model.
\end{abstract}

\bigskip

\begin{center}
Giorgio M. Vitetta, Emilio Sirignano and Francesco Montorsi

\vspace{5mm}University of Modena and Reggio Emilia

Department of Engineering "Enzo Ferrari"

Via P. Vivarelli 10/1, 41125 Modena - Italy

email: giorgio.vitetta@unimore.it, emilio.sirignano@unimore.it,
francesco.montorsi@gmail.com
\end{center}

\bigskip


\baselineskip0.2 in\newpage

\section{Introduction\label{sec:intro}}

\emph{Bayesian} \emph{filtering} and \emph{Bayesian} \emph{smoothing} for
\emph{state space models} (SSMs) are two interrelated problems that have
received significant attention for a number of years \cite{Anderson_1979}. Bayesian filtering allows to
recursively estimate, through a prediction/update mechanism, the
\emph{probability density function} (pdf) of the current state of any SSM,
given the history of some observed data up to the current time. Unluckily, the
general formulas describing the Bayesian filtering recursion (e.g., see
\cite[eqs. (4)-(5)]{Arulampalam_2002}) admit closed form solutions for
\emph{linear Gaussian} and \emph{linear Gaussian mixture} SSMs \cite{Anderson_1979} only. On the contrary, approximate solutions are available for
general \emph{nonlinear} models; these are based on \emph{sequential Monte
Carlo} (SMC) techniques (also known as \emph{particle filtering} methods)
which represent a powerful tool for numerical approximations
\cite{Doucet_2001}-\cite{Gustafsson_2010}.


Bayesian smoothing, instead, exploits an entire batch of measurements to
generate a significantly better estimate of the pdf (i.e., a \emph{smoothed}
or \emph{smoothing} pdf) of SSM state over a given observation interval. Two
general methods are available in the literature for recursively calculating
smoothing densities, namely the \emph{forward filtering-backward smoothing
recursion} \cite{Doucet_2000}, \cite{Kitagawa_1987} and the method based on
the \emph{two-filter smoothing formula} \cite{Kitagawa_1994}-\cite{Bresler_1986}.
 In both cases the computation of
smoothing densities requires combining the predicted and/or filtered densities
generated by a standard Bayesian filtering method with those produced by a
recursive backward technique (known as \emph{backward information filtering},
BIF, in the case of two-filter smoothing). Similarly as filtering, closed form
solutions for Bayesian smoothing are available for \emph{linear Gaussian} and
\emph{linear Gaussian mixture} models \cite{Anderson_1979}, \cite{Vo_2012}. This has motivated the
development of various SMC approximations (also known as \emph{particle
smoothers}) for the above mentioned two methods in the case of nonlinear SSMs
(e.g., see \cite{Doucet_2000}, \cite{Douc_2011}, \cite{Kitagawa_1994}, \cite{Kitagawa_1996}, \cite{Godsill_2004}-\cite{Lindsten_2013} and references therein).

While SMC methods can be directly applied to an arbitrary nonlinear SSM for
both filtering and smoothing, it has been recognized that their estimation
accuracy can be improved in the case of \emph{conditionally linear Gaussian}
(CLG) SSMs. In fact, the linear substructure of such models can be
marginalised, so reducing the dimension of their SMC space \cite{Chen_2000},
\cite{Schon_2005}. This idea has led to the development of important SMC
techniques for filtering and smoothing, known as \emph{Rao-Blackwellized}
\emph{particle filtering} (also dubbed \emph{marginalized particle filtering},
MPF) \cite{Schon_2005} and \emph{Rao-Blackwellized particle smoothing} (RBPS)
\cite{Briers_2010}, \cite{Fong_2002}, \cite{Lindsten_2016}, respectively.

Recently, the filtering problem for CLG\ SSMs has been investigated from a
\emph{factor graph} (FG) perspective in \cite{Vitetta_2016}, where a novel
interpretation of MPF as a \emph{forward only message passing algorithm} over
a specific FG has been provided and a novel extension of it, dubbed
\emph{turbo filtering} (TF), has been derived. In this manuscript, the same
conceptual approach is employed to provide new insights in the
\emph{fixed-interval} \emph{smoothing problem} \cite{Briers_2010} and to
develop a novel solution for it. The proposed solution is represented by a
novel RBPS method (dubbed \emph{Rao-Blackwellized serial smoothing}, RBSS)
having the following relevant features: a) it can be derived applying the well
known \emph{sum-product algorithm} (SPA) \cite{Loeliger_2007},
\cite{Kschischang_2001}, together with a specific scheduling procedure, to the
same FG\ developed in \cite{Vitetta_2016} for a CLG\ SSM; b) unlike the RBPS
methods devised in \cite{Briers_2010} and \cite{Fong_2002}, it can be employed
for a SSM in which both the linear and nonlinear state components influence
each another; c) its computational complexity is appreciably smaller than that
required by the other RBPS techniques; d) it benefits, unlike all the other
RBPS techniques, from the exploitation of all the available pseudo-measurements
and the \emph{ex novo} computation of the weights for the particles generated
in its forward recursion; e) it can be easily modified to compute the
\emph{joint smoothing distribution} over the entire observation interval (the
resulting algorithm is called \emph{extended} RBSS, ERBSS, in the following).
Our simulation results evidence that, for the considered SSM, RBSS 
achieves a good accuracy-complexity tradeoff and that, in particular, it is slightly outperformed by ERBSS in state estimation accuracy, which, however, at the price, however, of a substantially higher computational cost.

It is worth mentioning that the application of FG methods to Bayesian
smoothing is not new. However, as far as we know, the few results available in
the technical literature about this topic refer to the case of \emph{linear
Gaussian} SSMs only \cite{Loeliger_2007}, \cite{Loeliger_2016},
\cite{Wadehn_2016}, whereas we exclusively focus on the case in which the
mathematical laws expressing state dynamics and/or available observations are
\emph{nonlinear}.

The remaining part of this manuscript is organized as follows. The
 model of the considered CLG SSM is briefly illustrated in
Section \ref{sec:scenario}. A representation of the smoothing problem through
Forney-style FGs for both an arbitrary SSM and a CLG SSM is provided in Section
\ref{sec:Factorgraphs}. In Section \ref{sec:Message-Passing} the
RBSS technique is developed applying the SPA and proper message scheduling
strategies to the FG derived for a CLG SSM; moreover, it is shown how it can
be modified to estimate a point mass approximation of the joint smoothing
distribution. Our FG-based smoothing algorithms are compared, in terms of
accuracy and computational effort, in Section \ref{num_results}. Finally, some conclusions are offered in Section \ref{sec:conc}.

\emph{Notations}: The \emph{probability density function} (pdf) of a random
vector $\mathbf{R}$ evaluated at point $\mathbf{r}$ is denoted $f(\mathbf{r})
$; $\mathcal{N}\left(  \mathbf{r};\mathbf{\eta_{r}},\mathbf{C_{r}}\right)  $
represents the pdf of a Gaussian random vector $\mathbf{R}$ characterized by
the mean $\mathbf{\eta_{r}}$ and covariance matrix $\mathbf{\mathbf{C_{r}}}$
evaluated at point $\mathbf{r}$; the \emph{precision }(or \emph{weight})
\emph{matrix} associated with the covariance matrix $\mathbf{\mathbf{C_{r}}}$
is denoted $\mathbf{\mathbf{W_{r}}}$, whereas the \emph{transformed mean
vector} $\mathbf{\mathbf{W_{r}}\eta_{r}}$ is denoted $\mathbf{\mathbf{w_{r}}}%
$.

\section{System Model\label{sec:scenario}}

In the following we focus on the discrete-time CLG\ SSM described in
\cite{Vitetta_2016}, \cite{Vitetta_2017}. In brief, the SSM \emph{hidden state} in the $l$-th
interval is represented by the $D$-dimensional real vector $\mathbf{x}%
_{l}\triangleq\lbrack x_{0,l},x_{1,l},...,$ $x_{D-1,l}]^{T}$; this is
partitioned in a) its $D_{L}$-dimensional \emph{linear component }$\mathbf{x}_{l}^{(L)}\triangleq\lbrack x_{0,l}^{(L)},x_{1,l}^{(L)},...,x_{D_{L}%
-1,l}^{(L)}]^{T}$ and b) its $D_{N}$-dimensional \emph{nonlinear component
}$\mathbf{x}_{l}^{(N)}\triangleq\lbrack x_{0,l}^{(N)},x_{1,l}^{(N)}%
,...,x_{D_{N}-1,l}^{(L)}]^{T}$ (with $D_{L}<D$ and $D_{N}=D-D_{L}$). The
update equations of the linear and nonlinear components are given by
\begin{equation}
\mathbf{x}_{l+1}^{(L)}=\mathbf{A}_{l}^{(L)}\left(  \mathbf{x}_{l}%
^{(N)}\right)  \mathbf{x}_{l}^{(L)}+\mathbf{f}_{l}^{(L)}\left(  \mathbf{x}%
_{l}^{(N)}\right)  +\mathbf{w}_{l}^{(L)},\label{eq:XL_update}%
\end{equation}
and
\begin{equation}
\mathbf{x}_{l+1}^{(N)}=\mathbf{f}_{l}^{(N)}\left(  \mathbf{x}_{l}%
^{(N)}\right)  +\mathbf{A}_{l}^{(N)}\left(  \mathbf{x}_{l}^{(N)}\right)
\mathbf{x}_{l}^{(L)}+\mathbf{w}_{l}^{(N)}\text{,}\label{eq:XN_update}%
\end{equation}
respectively; here, $\mathbf{f}_{l}^{(L)}\left(\mathbf{x}\right)  $
($\mathbf{f}_{l}^{(N)}\left(\mathbf{x}\right)$) is a time-varying $D_{L}%
$-dimensional ($D_{N}$-dimensional) real function, $\mathbf{A}_{l}%
^{(L)}(\mathbf{x}_{l}^{(N)})$ ($\mathbf{A}_{l}^{(N)}(\mathbf{x}_{l}^{(N)})$)
is a time-varying $D_{L}\times D_{L}$ ($D_{N}\times D_{L}$) real matrix and
$\mathbf{w}_{l}^{(L)}$ ($\mathbf{w}_{l}^{(N)}$) is the $l$-th element of the
process noise sequence $\{\mathbf{w}_{k}^{(L)}\}$ ($\{\mathbf{w}_{k}^{(N)}%
\}$), which consists of $D_{L}$- dimensional ($D_{N}$-dimensional)
\emph{independent and identically distributed} (iid) noise\emph{\ }vectors
(statistical independence between $\{\mathbf{w}_{k}^{(L)}\}$ and
$\{\mathbf{w}_{k}^{(N)}\}$ is also assumed for simplicity). Moreover, in the
$l$-th interval some noisy observations, collected in the measurement vector
\begin{equation}
\mathbf{y}_{l}\triangleq\lbrack y_{0,l},y_{1,l},...,y_{P-1,l}]^{T}%
=\mathbf{h}_{l}\left(\mathbf{x}_{l}^{(N)}\right)  +\mathbf{B}_{l}\left(
\mathbf{x}_{l}^{(N)}\right)  \mathbf{x}_{l}^{(L)}+\mathbf{e}_{l}%
,\label{eq:y_t}%
\end{equation}
are available about $\mathbf{x}_{l}$; here, $\mathbf{B}_{l}(\mathbf{x}%
_{l}^{(N)})$ is a time-varying $P\times D_{L}$ real matrix, $\mathbf{h}%
_{l}(\mathbf{x}_{l}^{(N)})$ is a time-varying $P$-dimensional real function
and $\mathbf{e}_{l}$ the $l$-th element of the measurement noise sequence
$\left\{  \mathbf{e}_{k}\right\}  $ consisting of $P$-dimensional iid noise
vectors and independent of both $\{\mathbf{w}_{k}^{(N)}\}$ and $\{\mathbf{w}%
_{k}^{(L)}\}$. In the following Section we mainly focus on the so-called
\emph{fixed-interval} \emph{smoothing problem }\cite{Briers_2010}; this
consists of computing the sequence of \emph{posterior densities}%
$\{f(\mathbf{x}_{l}|\mathbf{y}_{1:N}),\,l=1,2,...,T\}$ (where $T$ represents
the length of the observation interval), given a) the \emph{initial} pdf
$f(\mathbf{x}_{1})$ and b) the $T\cdot P$-dimensional \emph{measurement}
vector $\mathbf{y}_{1:T}=\left[\mathbf{y}_{1}^{T},\mathbf{y}_{2}^{T},...,\mathbf{y}_{T}^{T}\right]^{T}$.

\section{A FG-Based Representation of the Smoothing
Problem\label{sec:Factorgraphs}}

In this Section we formulate the computation of the \emph{marginal smoothed
density} $f(\mathbf{x}_{l}|\mathbf{y}_{1:T})$ (with $l=1,2,...,T$) as a
message passing algorithm over a specific FG for the following two cases: C.1)
a SSM whose statistical behavior is characterized by the \emph{Markov
model} $f(\mathbf{x}_{l+1}|\mathbf{x}_{l})$ and the  
\emph{observation model} $f(\mathbf{y}_{l}|\mathbf{x}_{l})$; C.2) a SSM having the additional property of
being CLG (see the previous Section).

In case C.1 we take into consideration the \emph{joint} pdf $f(\mathbf{x}%
_{l},\mathbf{y}_{1:T})$ in place of the \emph{posterior} pdf
$f(\mathbf{x}_{l}|\mathbf{y}_{1:T})$. This choice is motivated by the fact
that: a) the computation of the former pdf can be easily formulated as a
\emph{recursive} message passing algorithm over a proper FG, since, as shown
below, this involves only products and sums of products; b) the former pdf,
being proportional to the latter one, is represented by the \emph{same} FG
(this issue is discussed in \cite[Sec. II, p. 1297]{Loeliger_2007}). Note that
the validity of statement a) relies on the following mathematical results: a) the factorization (e.g., see \cite[Sec. 3]{Kitagawa_1994})%
\begin{align}
f\left(\mathbf{x}_{l},\mathbf{y}_{1:T}\right)   &  =f\left(\mathbf{y}%
_{l:T}\left\vert \mathbf{x}_{l},\mathbf{y}_{1:(l-1)}\right.  \right)  f\left(
\mathbf{x}_{l},\mathbf{y}_{1:(l-1)}\right) \nonumber\\
&  =f\left(\mathbf{y}_{l:T}\left\vert \mathbf{x}_{l}\right.  \right)
f\left(\mathbf{x}_{l},\mathbf{y}_{1:(l-1)}\right) \label{factorisation}%
\end{align}
for the pdf of interest; b) the availability of recursive methods, known as \emph{Bayesian}
\emph{filtering} \cite{Arulampalam_2002} (and called \emph{forward filtering},
FF, in the following for clarity) and \emph{backward information filtering
}(BIF; e.g., see \cite{Kitagawa_1994}) for computing the joint pdf
$f(\mathbf{x}_{l},\mathbf{y}_{1:(l-1)})$ and the conditional pdf
$f(\mathbf{y}_{l:T}|\mathbf{x}_{l})$, respectively, for any $l$.

As far as FF is concerned, the formulation illustrated in \cite[Sec.
2]{Vitetta_2016} is adopted here; this consists of a \emph{measurement update}
(MU) step followed by a \emph{time update} (TU) step and assumes the a priori
knowledge of the pdf $f(\mathbf{x}_{1})$ for its initialization. In the MU
step of its $l$-th recursion (with $l=1,2,...,T$) the joint pdf
\begin{equation}
f\left(\mathbf{x}_{l},\mathbf{y}_{1:l}\right)  =f\left(\mathbf{x}%
_{l},\mathbf{y}_{1:(l-1)}\right)  f\left(  \mathbf{y}_{l}\left\vert
\mathbf{x}_{l}\right.  \right) \label{eq:meas_update}%
\end{equation}
is computed on the basis of pdf $f(\mathbf{x}_{l},\mathbf{y}_{1:(l-1)})$,
 and the new measurement vector $\mathbf{y}_{l}$. In the TU step, instead, 
 the pdf $f\left(\mathbf{x}_{l},\mathbf{y}_{1:l}\right)$ 
(\ref{eq:meas_update}) is exploited to compute the pdf
\begin{equation}
f\left(\mathbf{x}_{l+1},\mathbf{y}_{1:l}\right)  =\int f\left(
\mathbf{x}_{l+1}\left\vert \mathbf{x}_{l}\right.  \right)  f\left(
\mathbf{x}_{l},\mathbf{y}_{1:l}\right)  d\mathbf{x}_{l},\label{eq:time_update}%
\end{equation}
representing a \emph{prediction} about the future state $\mathbf{x}_{l+1}$.

A conceptually similar recursive procedure can be easily developed for the
$(T-l)$-th recursion of BIF (with $l=T-1,T-2,...,1$). In fact, this can be
formulated as a TU step followed by a MU step; these are expressed by
\begin{equation}
f\left(\mathbf{y}_{(l+1):T}\left\vert \mathbf{x}_{l}\right.  \right)  =\int
f\left(\mathbf{y}_{(l+1):T}\left\vert \mathbf{x}_{l+1}\right.  \right)
f\left(\mathbf{x}_{l+1}\left\vert \mathbf{x}_{l}\right.  \right)
d\mathbf{x}_{l}\label{eq:time_update_BF}%
\end{equation}
and
\begin{equation}
f\left(\mathbf{y}_{l:T}\left\vert \mathbf{x}_{l}\right.  \right)  =f\left(
\mathbf{y}_{(l+1):T}\left\vert \mathbf{x}_{l}\right.  \right)  f\left(
\mathbf{y}_{l}\left\vert \mathbf{x}_{l}\right.  \right)
,\label{eq:meas_update_BF}%
\end{equation}
respectively. Note that this procedure requires the knowledge of the pdf
$f(\mathbf{y}_{T}|\mathbf{x}_{T})$ for its initialization (see
(\ref{eq:time_update_BF})).

\begin{figure}[ptb]
\begin{center}
\includegraphics[scale = 0.7]{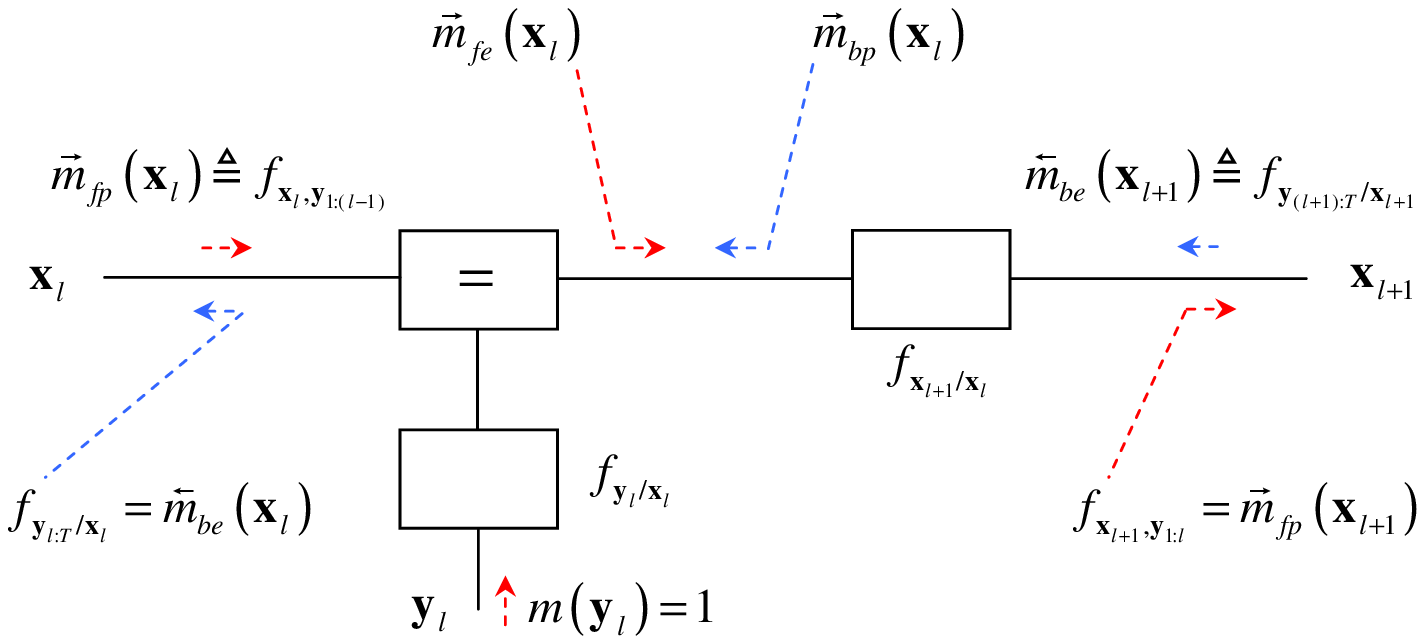}
\end{center}
\caption{Graphical representation of the message passing for the evaluation of
the joint pdf $f(\mathbf{x}_{l+1},\mathbf{y}_{1:l})$ and of the conditional
pdf $f(\mathbf{y}_{l:T}|\mathbf{x}_{l})$ on the basis of eqs.
(\ref{eq:meas_update})-(\ref{eq:time_update}) and (\ref{eq:time_update_BF}%
)-(\ref{eq:meas_update_BF}), respectively (the forward and backward message
flows are indicated by red and blue arrows, respectively)}%
\label{Fig_1}%
\end{figure}

Eqs. (\ref{eq:meas_update})-(\ref{eq:meas_update_BF}) show that each of the FF
(or BIF) recursions involves only \emph{products of pdfs} and a \emph{sum}
(i.e., an integration) \emph{of products}. For this reason, based on the
general rules about graphical models illustrated in \cite[Sect. II]%
{Loeliger_2007}, such recursions can be interpreted as specific instances of
the SPA\footnote{In a Forney-style FG, such a rule can be formulated as
follows \cite{Loeliger_2007}: the message emerging from a node \emph{f} along
some edge \emph{x} is formed as the product of \emph{f} and all the incoming
messages along all the edges that enter the node \emph{f} except \emph{x},
summed over all the involved variables except \emph{x}.} applied to the
\emph{cycle free} FG of Fig. \ref{Fig_1} (where the simplified notation of 
\cite{Loeliger_2007} is employed). 
More specifically, it is easy to show
that eqs. (\ref{eq:meas_update}) and (\ref{eq:time_update}) can be seen as a
SPA-based algorithm for \emph{forward} message passing over 
the FG shown in Fig. \ref{Fig_1} (the flow of
forward messages is indicated by \emph{red} arrows in the considered figure). In
fact, if the FG is fed by the message\footnote{In the following the acronyms
\emph{be}, \emph{fp} and \emph{sm} are employed in the subscripts
of various messages, so that readers can easily understand their meaning; in
fact, the messages these acronyms refer to represent a form of \emph{backward
estimation}, \emph{forward prediction} and
\emph{smoothing}, respectively.}%
\begin{equation}
\vec{m}_{fp}\left(\mathbf{x}_{l}\right)  \triangleq f(\mathbf{x}%
_{l},\mathbf{y}_{1:(l-1)}),\label{input_forwb}%
\end{equation}
the forward messages emerging from the \emph{equality node} and that passed
along the edge associated with $\mathbf{x}_{l+1}$ are given by $\vec{m}%
_{fe}\left(\mathbf{x}_{l}\right)  =f\left(\mathbf{x}_{l},\mathbf{y}%
_{1:l}\right)  $ and $f(\mathbf{x}_{l+1},\mathbf{y}_{1:l})=\vec{m}_{fp}\left(
\mathbf{x}_{l+1}\right)  $, respectively \cite{Vitetta_2016}, \cite{Vitetta_2017}. A similar
interpretation can be provided for eqs. (\ref{eq:time_update_BF}) and
(\ref{eq:meas_update_BF}), which, however, can be reformulated as a SPA-based
algorithm for \emph{backward} message passing over the considered FG. In fact, if the input message%
\begin{equation}
\overset{\leftarrow}{m}_{be}\left(\mathbf{x}_{l+1}\right)  \triangleq
f\left( \mathbf{y}_{(l+1):T}\left\vert \mathbf{x}_{l+1}\right.  \right)
\label{input_back}%
\end{equation}
enters the FG along the half edge associated with $\mathbf{x}_{l+1}$ (the flow
of backward messages is indicated by \emph{blue} arrows in Fig. \ref{Fig_1}), the
backward message $\overset{\leftarrow}{m}_{bp}\left(  \mathbf{x}_{l}\right)  $
\ emerging from the node associated with the pdf $f(\mathbf{x}_{l+1}%
|\mathbf{x}_{l})$ is given by (see (\ref{eq:time_update_BF}))%
\begin{align}
\overset{\leftarrow}{m}_{bp}\left(\mathbf{x}_{l}\right)   &  =\int
\overset{\leftarrow}{m}_{be}\left(\mathbf{x}_{l}\right)  f\left(
\mathbf{x}_{l+1}\left\vert \mathbf{x}_{l}\right.  \right)  d\mathbf{x}%
_{l}\nonumber\\
&  =\int f(\mathbf{y}_{(l+1):T}|\mathbf{x}_{l+1})f\left(\mathbf{x}%
_{l+1}\left\vert \mathbf{x}_{l}\right.  \right)  d\mathbf{x}_{l}\nonumber\\
&  =f\left(\mathbf{y}_{(l+1):T}\left\vert \mathbf{x}_{l}\right.  \right)
\text{.}\label{eq:mess_bp}%
\end{align}
Therefore, the message going out of the equality node in the backward
direction can be evaluated as (see (\ref{eq:meas_update_BF}) and
(\ref{input_back}))
\begin{align}
&  f\left(\mathbf{y}_{l}\left\vert \mathbf{x}_{l}\right.  \right)
\overset{\leftarrow}{m}_{bp}\left(\mathbf{x}_{l}\right) = f\left(\mathbf{y}_{l}\left\vert \mathbf{x}_{l}\right.  \right)  f\left(
\mathbf{y}_{(l+1):T}\left\vert \mathbf{x}_{l}\right.  \right) \nonumber\\
&  =f\left(  \mathbf{y}_{l:T}\left\vert \mathbf{x}_{l}\right.  \right)
=\overset{\leftarrow}{m}_{be}\left(\mathbf{x}_{l}\right)
\label{eq:mess_eq_back}%
\end{align}
and this concludes our proof.

These results easily lead to the conclusion that, once the forward and
backward message passing algorithms illustrated above have been carried out
over the entire observation interval, the smoothed pdf $f\left(
\mathbf{x}_{l},\mathbf{y}_{1:T}\right)$ can be evaluated as (see
(\ref{factorisation}), (\ref{input_forwb}) and (\ref{eq:mess_eq_back}))%
\begin{equation}
f\left(\mathbf{x}_{l},\mathbf{y}_{1:T}\right)  =\vec{m}_{fp}\left(
\mathbf{x}_{l}\right)  \overset{\leftarrow}{m}_{be}\left(\mathbf{x}%
_{l}\right)  \text{,}\label{eq:smooth_dens_1}%
\end{equation}
with $l=1,2,...,T$ (note that $\overset{\leftarrow}{m}_{be}\left(
\mathbf{x}_{T}\right) =1$ and $\vec{m}_{fp}\left(\mathbf{x}_{1}\right)  =f(\mathbf{x}_{1})$) 

\begin{figure}[ptb]
\vspace{5mm}
\begin{center}
\includegraphics[scale = 0.6]{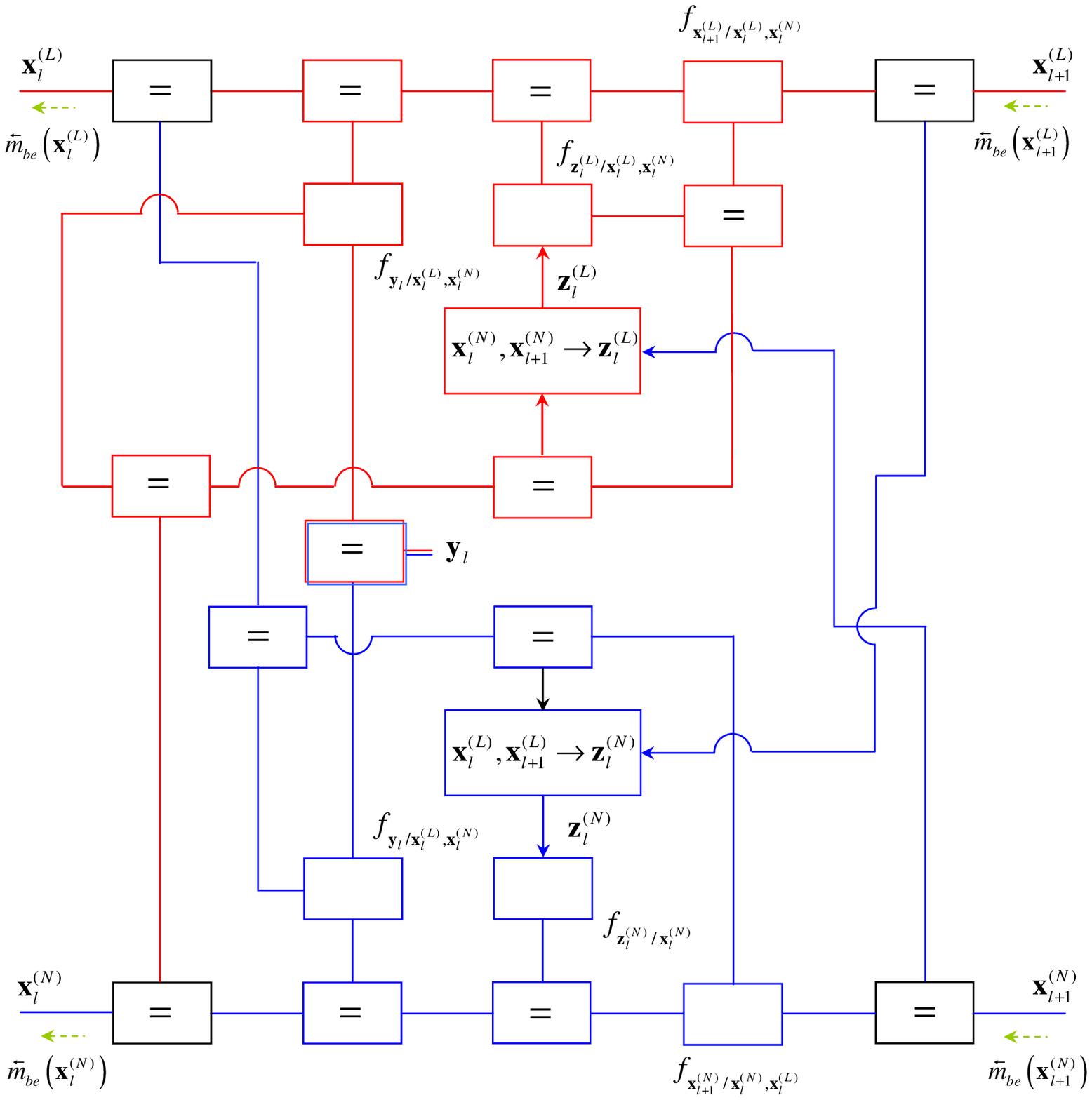}
\end{center}
\caption{Factor graph resulting from the merge of two sub-graphs, one referring to the smoothing problem for the linear state component, the other one to that for the nonlinear state component (these are identified by red and blue lines, respectively, whereas the equality nodes introduced to merge them by black lines). The direction of the messages passed over the half edges $\mathbf{x}_{l}^{(L)}$ and $\mathbf{x}_{l}^{(N)}$ (inputs) and over the half edges $\mathbf{x}_{l+1}^{(L)}$ and $\mathbf{x}_{l+1}^{(N)}$ (outputs) is indicated by green arrows.}%
\label{Fig_2}%
\end{figure}

The FG we develop for case C.2 is based not only on that analysed for case C.1,
but also on the idea of representing a mixed linear/nonlinear SSM
as the concatenation of two interacting sub-models, one referring to the
linear component of system state, the other one to its nonlinear
component \cite{Vitetta_2016}. This suggests to \emph{decouple} the smoothing problem for
$\mathbf{x}_{l}^{(L)}$ from that for $\mathbf{x}_{l}^{(N)}$, i.e. the
evaluation of \emph{\ }$f(\mathbf{x}_{l}^{(L)}|\mathbf{y}_{1:T})$ from that of
$f(\mathbf{x}_{l}^{(N)}|\mathbf{y}_{1:T})$. In practice, from a graphical
viewpoint, two sub-graphs, one referring to smoothing for $\mathbf{x}%
_{l}^{(L)}$, the other one to smoothing for $\mathbf{x}_{l}^{(N)}$, are
developed first; then, they are merged by adding five distinct equality nodes,
associated with the variables (namely, $\mathbf{y}_{l}$, $\mathbf{x}_{l}^{(L)}$, $\mathbf{x}%
_{l}^{(N)}$, $\mathbf{x}_{l+1}^{(L)}$ and $\mathbf{x}_{l+1}^{(N)}$) shared by
such sub-graphs. This leads to the FG\ illustrated in Fig. \ref{Fig_2}, in
which the sub-graph referring to the linear (nonlinear) state component is 
identified by red (blue) lines, whereas the
equality nodes added to merge them are identified by black lines. Note that 
 the sub-graph for the linear (nonlinear)
component is derived under the assumption that the nonlinear (linear)
component is \emph{known}. Consequently, smoothing for the linear component
$\mathbf{x}_{l}^{(L)}$ can benefit not only from the measurement
$\mathbf{y}_{l}$, but also from the so called \emph{pseudo-measurement} (see
(\ref{eq:XN_update}))
\begin{equation}
\mathbf{z}_{l}^{(L)}\triangleq\mathbf{x}_{l+1}^{(N)}-\mathbf{f}_{l}%
^{(N)}\left(  \mathbf{x}_{l}^{(N)}\right)  =\mathbf{A}_{l}^{(N)}\left(
\mathbf{x}_{l}^{(N)}\right)  \mathbf{x}_{l}^{(L)}+\mathbf{w}_{l}%
^{(N)},\label{eq:z_L_l}%
\end{equation}
which, from a statistical viewpoint, is characterized by the pdf $f(\mathbf{z}%
_{l}^{(L)}|\mathbf{x}_{l}^{(L)},\mathbf{x}_{l}^{(N)})$. Similarly, the
pseudo-measurement (see (\ref{eq:XL_update}))
\begin{equation}
\mathbf{z}_{l}^{(N)}\triangleq\mathbf{x}_{l+1}^{(L)}-\mathbf{A}_{l}%
^{(L)}\left(  \mathbf{x}_{l}^{(N)}\right)  \mathbf{x}_{l}^{(L)}=\mathbf{f}%
_{l}^{(L)}\left(  \mathbf{x}_{l}^{(N)}\right)  +\mathbf{w}_{l}^{(L)}%
\text{,}\label{eq:z_N_l}%
\end{equation}
characterized by the pdf $f(\mathbf{z}_{l}^{(N)}|\mathbf{x}_{l}^{(N)})$, can
be exploited in smoothing for the nonlinear component
$\mathbf{x}_{l}^{(N)}$. \ These considerations explain why the upper (lower)
sub-graph shown in Fig. \ref{Fig_2} contains an additional node representing
the pdf $f(\mathbf{z}_{l}^{(L)}|\mathbf{x}_{l}^{(L)},\mathbf{x}_{l}^{(N)})$
($f(\mathbf{z}_{l}^{(N)}|\mathbf{x}_{l}^{(N)})$) and a \emph{specific}
\emph{node} not referring to the above mentioned pdf factorizations, but
representing the \emph{transformation} from the couple $(\mathbf{x}_{l}%
^{(N)},\mathbf{x}_{l+1}^{(N)})$ to $\mathbf{z}_{l}^{(L)}$ ($(\mathbf{x}%
_{l}^{(L)},\mathbf{x}_{l+1}^{(L)})$ to $\mathbf{z}_{l}^{(N)}$); the last
peculiarity, evidenced by the presence of an arrow on all the edges connected
to such a node, has to be carefully kept into account when deriving message
passing algorithms.

Given the FG of Fig. \ref{Fig_2}, we would like to follow the same line of
reasoning as that illustrated for the graphical model of Fig. \ref{Fig_1}. In
particular, given the input backward messages $\overset{\leftarrow}{m}%
_{be}(\mathbf{x}_{l+1}^{(L)})\triangleq f(\mathbf{y}_{(l+1):T},\mathbf{z}%
_{(l+1):T}^{(L)},\mathbf{x}_{l+1}^{(L)})$ and $\overset{\leftarrow}{m}%
_{be}(\mathbf{x}_{l+1}^{(N)})\triangleq f(\mathbf{y}_{(l+1):T},\mathbf{z}%
_{(l+1):T}^{(N)},\mathbf{x}_{l+1}^{(N)})$, we would like to derive a BIF algorithm based on this FG
 (FF has already been investigated in \cite{Vitetta_2016} and \cite{Vitetta_2017}) and generating the
output backward messages $\overset{\leftarrow}{m}_{be}(\mathbf{x}_{l}%
^{(L)})=f(\mathbf{y}_{l:T},\mathbf{z}_{l:T}^{(L)},\mathbf{x}_{l}^{(L)})$ and
$\overset{\leftarrow}{m}_{be}(\mathbf{x}_{l}^{(N)})=f(\mathbf{y}%
_{l:T},\mathbf{z}_{l:T}^{(N)},\mathbf{x}_{l}^{(N)})$ on the basis of the available a priori information
and the noisy measurement $\mathbf{y}_{l}$. Unluckily, the new FG, unlike the
one represented in Fig. \ref{Fig_1}, is \emph{not cycle-free}, so that any
application of the SPA to it unavoidably leads to\emph{\ approximate
solutions} \cite{Kschischang_2001}, whatever \emph{message scheduling
procedure} is adopted. In the following Section we show that the RBSS
technique we propose represents one of such solutions.

\section{Particle Smoothing as Message Passing\label{sec:Message-Passing}}

In this Section we first illustrate some assumptions about the statistical
properties of the SSM defined in Section \ref{sec:scenario}. Then, we develop
the RBSS technique and compare its most relevant features 
with those of the other RBPS algorithms available in the
technical literature. Finally, we show how this technique can be modified to
estimate the \emph{joint smoothing density }$f(\mathbf{x}_{1:T}|\mathbf{y}%
_{1:T})$.

\subsection{Statistical properties of the considered SSM\label{Stat_assump}}

Even if the FG\ representation shown in Fig. \ref{Fig_2} can be employed for
any mixed linear/nonlinear system described by eqs. (\ref{eq:XL_update}%
)-(\ref{eq:y_t}), the methods derived in this Section apply, like MPF
\cite{Schon_2005} and TF \cite{Vitetta_2016}, to the specific class of
GLG\emph{\ }SSMs. For this reason, following \cite{Vitetta_2016}, \cite{Vitetta_2017} we assume
that: a) the process noise $\{\mathbf{w}_{k}^{(L)}\}$ ($\{\mathbf{w}_{k}%
^{(N)}\}$) is Gaussian and all its elements have zero mean
and covariance $\mathbf{C}_{w}^{(L)}$ ($\mathbf{C}_{w}^{(N)}$) for any $l$; b)
the measurement noise$\ \{\mathbf{e}_{k}^{(L)}\}$ is Gaussian
having zero mean and covariance matrix $\mathbf{C}_{e}$ for any $l$; c) all
the above mentioned Gaussian processes are statistically independent. Under
these assumptions, the pdfs $f(\mathbf{y}_{l}|\mathbf{x}_{l}^{(L)}%
,\mathbf{x}_{l}^{(N})$, $f(\mathbf{z}_{l}^{(L)}|\mathbf{x}_{l}^{(L)})$ and
$f(\mathbf{x}_{l+1}^{(L)}|\mathbf{x}_{l}^{(L)},\mathbf{x}_{l}^{(N)})$ are Gaussian with mean (covariance matrix) $\mathbf{B}_{l}(\mathbf{x}_{l}^{(N)}%
)\mathbf{x}_{l}^{(L)}+\mathbf{h}_{l}(\mathbf{x}_{l}^{(N)})$, $\mathbf{A}%
_{l}^{(N)}(\mathbf{x}_{l}^{(N)})\,\mathbf{x}_{l}^{(L)}$ and $\mathbf{f}%
_{l}^{(L)}(\mathbf{x}_{l}^{(N)})+\mathbf{A}_{l}^{(L)}(\mathbf{x}_{l}%
^{(N)})\,\mathbf{x}_{l}^{(L)}$, respectively ($\mathbf{C}_{e}$, $\mathbf{C}%
_{w}^{(N)}$ and $\mathbf{C}_{w}^{(L)}$, respectively). Similarly, the pdfs
$f(\mathbf{z}_{l}^{(N)}|\mathbf{x}_{l}^{(N)})$ and $f(\mathbf{x}_{l+1}%
^{(N)}|\mathbf{x}_{l}^{(N)},\mathbf{x}_{l}^{(L)})$ are
Gaussian with mean (covariance matrix) $\mathbf{f}_{l}^{(L)}(\mathbf{x}%
_{l}^{(N)})$ and $\mathbf{f}_{l}^{(N)}(\mathbf{x}_{l}^{(N)})+\mathbf{A}%
_{l}^{(N)}(\mathbf{x}_{l}^{(N)})\,\mathbf{x}_{l}^{(L)}$, respectively
($\mathbf{C}_{w}^{(L)}$ and $\mathbf{C}_{w}^{(N)}$, respectively).

\subsection{Derivation of the Rao-Blacwellized serial
smoother\label{RBSS_derivation}}

The FF algorithm employed in the \emph{forward pass} of the proposed RBSS is
represented by MPF\footnote{Note that TF can be employed in place of MPF in the forward pass of RBSS. However, our computer simulations have evidenced that, in the presence
of strong measurement and/or process noise (like in the scenarios considered
in Section \ref{num_results}), this choice doe not provide any performance
improvement with respect to MPF.}. In its $(l-1)$-th recursion (with
$l=2,3,...,T$), the particle set $\{\mathbf{x}_{l/(l-1),j}^{(N)}%
,j=0,1,...,N_{p}-1\}$, consisting of $N_{p}$ distinct particles, is
\emph{predicted} for the \emph{nonlinear state component} $\mathbf{x}_{l}^{(N)}$
(TU for this component); the weight $w_{l/(l-1),j}$ assigned to the particle
$\mathbf{x}_{l/(l-1),j}^{(N)}$ is equal to $1/N_{p}$ for any $j$, since the use of particle resampling in each recursion is assumed. The particle
weights are updated in the MU of the following (i.e., $l$-th) recursion on the
basis of the new measurement $\mathbf{y}_{l}$ (MU for the nonlinear
component): the new weights are denoted $\{w_{l/l,j},j=0,1,...,N_{p}-1\}$ in
the following and, generally speaking, are all different. This is followed by
particle resampling, that generates the new particle set $\{\mathbf{x}%
_{l/l,j}^{(N)},j=0,1,...,N_{p}-1\}$ (usually containing multiple copies of
the most likely particles of the set $\{\mathbf{x}_{l/(l-1),j}^{(N)}\}$). 
A conceptually similar procedure is followed for
the \emph{linear state component}, for which a particle-dependent Gaussian
representation is adopted. In particular, in the following, the Gaussian model
predicted for $\mathbf{x}_{l}^{(L)}$ in the $(l-1)$-th recursion (TU\ for the
linear state component) and associated with $\mathbf{x}_{l/(l-1),j}^{(N)}$ is
denoted $\mathcal{N}(\mathbf{x}_{l}^{(L)};\mathbf{\eta}_{fp,l,j}%
^{(L)},\mathbf{C}_{fp,l,j}^{(L)})$. Note that only a portion of these Gaussian
models is usually updated in the MU of the next (i.e., $l$-th) recursion; in
fact, this task follows particle resampling, which typically leads to
discarding a fraction of the particles collected in the set $\{\mathbf{x}%
_{l/(l-1),j}^{(N)}\}$.

\begin{figure}[ptb]
\begin{center}
\includegraphics[scale = 0.6]{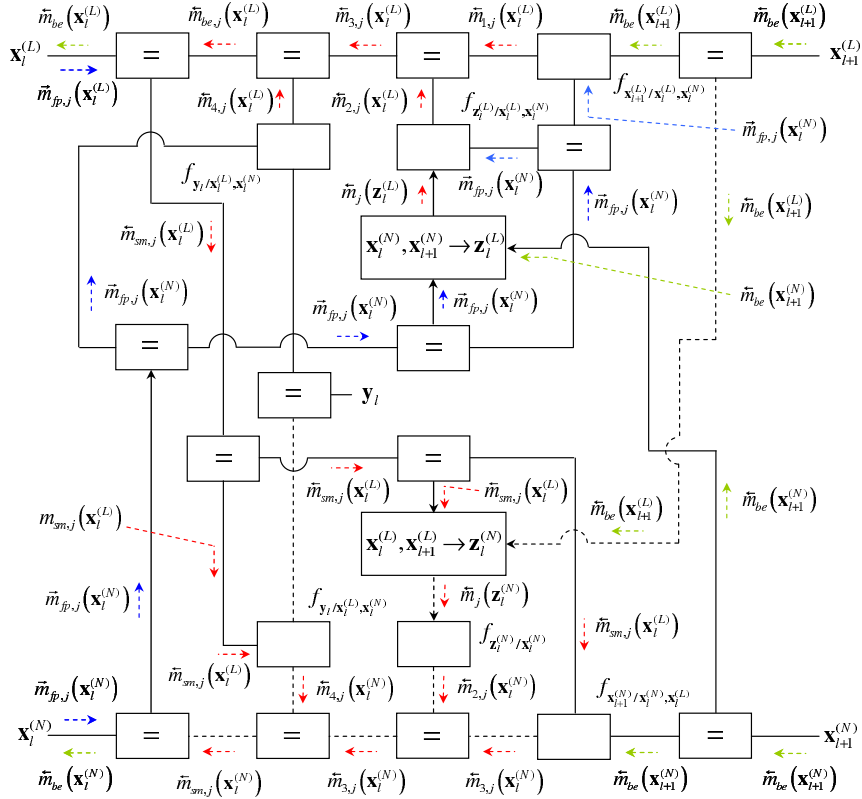}
\end{center}
\vspace{10mm}
\caption{Representation of the message scheduling employed in the $(T-l)$-th
recursion of RBSS backward processing. The edges involved in the first
(second) part of message passing are identified by continuous (dashed) lines.
Blue, green and red arrows are employed to identify the input forward
messages, the input/output backward messages and the remaining messages,
respectively.}%
\label{Fig_4}%
\end{figure}

The recursive algorithm developed for the \emph{backward pass} of the RBSS
technique results from the application of the SPA\ to the FG\ shown in Fig.
\ref{Fig_2}, and accomplishes BIF and smoothing (i.e., the merge of statistical
information generated by FF and BIF). Each of its recursions consists of two
parts, the first concerning the linear state component, the second one the
nonlinear state component; moreover, these parts are executed \emph{serially}.
The \emph{message scheduling} employed in the $(T-l)$-th recursion of BIF and
smoothing (with $l=T-1,T-2,...,1$) is summarized in Fig. \ref{Fig_4}, where
the edges involved in the first (second) part are identified by continuous
(dashed) lines. Similarly to MPF, most of the processing tasks which both parts
consist of can be formulated with reference to a single particle; this
explains why the notation adopted for the messages appearing in Fig.
\ref{Fig_4} includes the subscript $j$, that represents the index of the
particle (namely, the particle $\mathbf{x}_{l/(l-1),j}^{(N)}$) representing
$\mathbf{x}_{l}^{(N)}$ within the considered recursion.%

Before providing a detailed description of the messages passed in the
graphical model of Fig. \ref{Fig_4}, all the messages
feeding the considered recursion (i.e., its \emph{input} \emph{messages}) and
those emerging from it (i.e., its \emph{output messages}) must be defined. The
\emph{input messages} can be divided in two groups. The first group consists
of the messages $\vec{m}_{fp,j}(\mathbf{x}_{l}^{(L)})$ and $\vec{m}%
_{fp,j}(\mathbf{x}_{l}^{(N)})$, that are predicted the $(l-1)$-th recursion
of the forward pass; the second one, instead, is made of the messages
$\overset{\leftarrow}{m}_{be,j}(\mathbf{x}_{l+1}^{(N)})$ and $\overset
{\leftarrow}{m}_{be,j}(\mathbf{x}_{l+1}^{(L)})$, that are generated in the
$(T-l-1)$-th recursion of the backward pass. The messages of the \emph{first
group} are defined as
\begin{equation}
\vec{m}_{fp,j}\left(  \mathbf{x}_{l}^{(N)}\right)  \triangleq\delta\left(
\mathbf{x}_{l}^{(N)}-\mathbf{x}_{l/(l-1),j}^{(N)}\right) \label{m_fp_N}%
\end{equation}
and%
\begin{equation}
\vec{m}_{fp,j}\left(\mathbf{x}_{l}^{(L)}\right)  \triangleq\mathcal{N}%
\left(\mathbf{x}_{l}^{(L)};\mathbf{\eta}_{fp,l,j}^{(L)},\mathbf{C}%
_{fp,l,j}^{(L)}\right),\label{m_fp_L}%
\end{equation}
and can be interpreted as
the $j$-th \emph{hypothesis} about a) the value (namely, $\mathbf{x}%
_{l/(l-1),j}^{(N)}$) taken on by the (hidden) nonlinear state component
$\mathbf{x}_{l}^{(N)}$ and b) the statistical representation of the (hidden)
linear state component $\mathbf{x}_{l}^{(L)}$ associated with such a value, respectively. In
the $l$-th recursion of FF, the likelihood of this hypothesis is assessed by
evaluating the above mentioned weight $w_{l/l,j}$; such a weight, however, is
\emph{ignored} in the backward pass. This choice is motivated by the our belief that, if such
a weight is computed \emph{ex novo}, its accuracy can be improved thanks to
the availability of both more refined (i.e., smoothed) statistical information
about $\mathbf{x}_{l}^{(L)}$ and additional (backward) information about
$\mathbf{x}_{l+1}^{(N)}$ (see (\ref{m_be_N}) and (\ref{m_be_L}) below).

The input messages of the \emph{second group} are defined as
\begin{equation}
\overset{\leftarrow}{m}_{be}\left(  \mathbf{x}_{l+1}^{(N)}\right)
\triangleq\delta\left(  \mathbf{x}_{l+1}^{(N)}-\mathbf{x}_{be,l+1}%
^{(N)}\right) \label{m_be_N}%
\end{equation}
and
\begin{equation}
\overset{\leftarrow}{m}_{be}\left(  \mathbf{x}_{l+1}^{(L)}\right)
\triangleq\mathcal{N}\left(  \mathbf{x}_{l+1}^{(L)};\mathbf{\eta}%
_{be,l+1}^{(L)},\mathbf{C}_{be,l+1}^{(L)}\right),\label{m_be_L}%
\end{equation}
and represent part of the statistical information generated in the previous
(i.e., the $(T-l-1)$-th) recursion of the backward pass. In particular, as
explained in detail below, the messages $\overset{\leftarrow}{m}%
_{be}(\mathbf{x}_{l+1}^{(N)})$ and $\overset{\leftarrow}{m}_{be}%
(\mathbf{x}_{l+1}^{(L)})$ convey the \emph{final estimate} $\mathbf{x}%
_{be,l+1}^{(N)}$ (i.e., a single particle representation) of $\mathbf{x}%
_{l+1}^{(N)}$ and a \emph{simplified statistical representation} of
$\mathbf{x}_{l+1}^{(L)}$, respectively. This explains why the RBSS, in the
$(T-l)$-th recursion of its backward pass, processes the input messages
(\ref{m_fp_N})-(\ref{m_be_L}) to compute an estimate, denoted $\mathbf{x}%
_{be,l}^{(N)}$, of $\mathbf{x}_{l}^{(N)}$ and a simplified statistical model,
denoted $\mathcal{N}(\mathbf{x}_{l}^{(L)};\mathbf{\eta}_{be,l}^{(L)}%
,\mathbf{C}_{be,l}^{(L)})$, for $\mathbf{x}_{l}^{(L)}$; these information are
conveyed by the \emph{output messages} $\overset{\leftarrow}{m}_{be}%
(\mathbf{x}_{l}^{(N)})$ and $\overset{\leftarrow}{m}_{be}(\mathbf{x}_{l}%
^{(L)})$, respectively. The evaluation of these messages is based, as already
mentioned above, on the scheduling illustrated in Fig. \ref{Fig_4} and on the
formulas listed in Tables \ref{Table1} and
\ref{Table2} (actually, the only formulas missing in these Tables are those
employed in the evaluation of the message $\overset{\leftarrow}{m}%
_{j}(\mathbf{z}_{l}^{(N)}) $ (\ref{m_z_N}) \ and, in particular, of its
parameters $\mathbf{\eta}_{\mathbf{z},l,j}^{(N)}$ (\ref{eta_z_N}) and
$\mathbf{C}_{\mathbf{z},l,j}^{(N)} $ (\ref{C_z_N}); mathematical details about
this can be found in \cite[Sec. 6]{Vitetta_2016}). Such formulas refer to 
the computation of the message $m_{out}\left(\mathbf{x}\right)  = m_{in,1}\left(\mathbf{x}\right)
m_{in,2}\left(\mathbf{x}\right)$ (emerging
from an \emph{equality node} fed by the messages $m_{in,1}\left(
\mathbf{x}\right)  $ and $m_{in,2}\left(  \mathbf{x}\right)  $)
and 
\begin{equation}
m_{out}\left(  \mathbf{x}_{2}\right)  =%
{\displaystyle\int}
m_{in}\left(  \mathbf{x}_{1}\right)  f\left(  \mathbf{x}_{1},\mathbf{x}%
_{2}\right)  d\mathbf{x}_{1}\label{formula2}%
\end{equation}
(emerging from a \emph{function node} $f\left(\mathbf{x}_{1}%
,\mathbf{x}_{2}\right)  $ fed by the message $m_{in,}\left(\mathbf{x}%
_{1}\right)$), respectively; moreover, they are provided by \cite[Table 2, p. 1303]{Loeliger_2007} or can be easily derived on the basis of standard mathematical results about
Gaussian random variables. For this
reason, in the following description of the RBSS backward pass, we provide,
for each message, a simple code identifying the specific formula on which its
evaluation is based; in particular, the notation TX-Y is employed to identify
formula no. Y appearing in Table X. Moreover, to ease the interpretation of
the proposed signal processing tasks executed within the RBSS algorithm, the
message passing accomplished in the considered recursion is divided in the
seven steps described below; steps 1-3 and steps 4-6 refer to the
two parts of the message passing shown in Fig. \ref{Fig_4}, whereas the last step concern the evaluation of: a) the smoothed
pdf of $\mathbf{x}_{l}$ and the pdfs of its components; b) the output messages
$\overset{\leftarrow}{m}_{be}(\mathbf{x}_{l}^{(N)})$ and $\overset{\leftarrow
}{m}_{be}(\mathbf{x}_{l}^{(L)})$.%

\begin{table*}
\centering%
\begin{tabular}
[c]{|llll|}\hline\hline
Formula no. & $m_{in,1}(\mathbf{x})$ & $m_{in,2}(\mathbf{x})$ & $m_{out}%
(\mathbf{x})$\\\hline\hline
1 & $\delta\left( \mathbf{x}-\mathbf{a}\right)  $ & $f(\mathbf{x})$ &
$f(\mathbf{a})\,\delta\left(  \mathbf{x}-\mathbf{a}\right)  $\\\hline
2 & $\mathcal{N}\left(  \mathbf{x};\mathbf{\eta}_{1},\mathbf{C}_{1}\right)  $
& $\mathcal{N}\left(  \mathbf{x};\mathbf{\eta}_{2},\mathbf{C}_{2}\right)  $ &
\begin{tabular}
[c]{l}%
$\mathcal{N}\left(  \mathbf{x};\mathbf{\eta},\mathbf{C}\right),$\\
$\mathbf{w}=\mathbf{w}_{1}+\mathbf{w}_{2}$, $\mathbf{W}=\mathbf{W}_{1}+\mathbf{W}_{2}$%
\end{tabular}
\\\hline
3 & $\mathcal{N}\left(  \mathbf{x};\mathbf{\eta}_{1},\mathbf{C}_{1}\right)  $
& $\mathcal{\mathcal{N}}\left(  \mathbf{c};\mathbf{Ax+b},\mathbf{C}%
_{2}\right)  $ &
\begin{tabular}
[c]{l}%
$\mathcal{N}\left( \mathbf{x};\mathbf{\eta},\mathbf{C}\right)  ,$\\
$\mathbf{w}=\mathbf{w}_{1}+\mathbf{A}^{T}\mathbf{W}_{2}\ \mathbf{c}$, $\mathbf{W}=\mathbf{W}_{1}+\mathbf{A}^{T}\mathbf{W}_{2}\mathbf{A}$%
\end{tabular}
\\\hline
\end{tabular}
\caption{Mathematical rules for the evaluation of the message $m_{out}(\mathbf{x})$,
emerging from an equality node fed by the input messages $m_{in,1}(\mathbf{x})$ and $m_{in,2}(\mathbf{x})$.
\label{Table1}}%
\end{table*}

\begin{table*}[tbp]%
\centering%
\begin{tabular}
[c]{|llll|}\hline\hline
Formula no. & $m_{in}(\mathbf{x}_{1})$ & $f(\mathbf{x}_{1},\mathbf{x}_{2})$ &
$m_{out}(\mathbf{x}_{2})$\\\hline\hline
1 & $\mathcal{N}\left(  \mathbf{x}_{1};\mathbf{\eta}_{1},\mathbf{C}%
_{1}\right)  $ & $\mathcal{N}\left(\mathbf{x}_{2};\mathbf{Ax}_{1}%
+\mathbf{g},\mathbf{C}_{2}\right)  $ & $\mathcal{N}\left(  \mathbf{x}%
_{2};\mathbf{A\eta}_{1}+\mathbf{g},\mathbf{AC}_{1}\mathbf{A}_{l}%
^{T}+\mathbf{C}_{2}\right)  $\\\hline
2 & $\delta\left(  \mathbf{x}_{1}-\mathbf{a}\right)  $ & $\mathcal{N}\left(
\mathbf{x}_{2};\mathbf{Ax}_{1}+\mathbf{g},\mathbf{C}_{2}\right)  $ &
$\mathcal{N}\left(\mathbf{x}_{2};\mathbf{Aa}+\mathbf{g},\mathbf{C}%
_{2}\right)  $\\\hline
3 & $\delta\left(\mathbf{x}_{1}-\mathbf{a}\right)  $ & $\mathcal{N}\left(
\mathbf{x}_{1};\mathbf{Ax}_{2},\mathbf{C}_{2}\right)  $ & $\mathcal{N}\left(
\mathbf{a};\mathbf{Ax}_{2},\mathbf{C}_{2}\right)  $\\\hline
4 & $\mathcal{N}\left(  \mathbf{x}_{1};\mathbf{\eta}_{1},\mathbf{C}%
_{1}\right)  $ & $\mathcal{N}\left(  \mathbf{x}_{1};\mathbf{\eta}%
_{2},\mathbf{C}_{2}\right)  $ & $%
\begin{array}
[c]{c}%
K\exp\left\{ \frac{1}{2}\left[  \mathbf{\eta}^{T}\mathbf{W\eta}-\mathbf{\eta
}_{1}^{T}\mathbf{W}_{1}\mathbf{\eta}_{1}-\mathbf{\eta}_{2}^{T}\mathbf{W}%
_{2}\mathbf{\eta}_{2}\right]  \right\} \\
\mathbf{w}=\mathbf{W}_{1}\mathbf{\eta}_{1}+\mathbf{W}_{2}\mathbf{\eta}_{2},\mathbf{W}=\mathbf{W}_{1}+\mathbf{W}_{2},\\
K=(\det(\mathbf{C}_{1}+\mathbf{C}_{2}))^{-N/2}%
\end{array}
$\\\hline
5 & $\mathcal{N}\left(\mathbf{x}_{1};\mathbf{\eta}_{1},\mathbf{C}%
_{1}\right)  $ & $\mathcal{N}\left(  \mathbf{x}_{1};\mathbf{g}+\mathbf{A}%
\,\mathbf{x}_{2},\mathbf{C}_{2}\right)  $ & $%
\begin{array}
[c]{c}%
\mathcal{N}\left(\mathbf{x}_{2};\mathbf{\eta},\mathbf{C}\right) \\
\mathbf{w}=\mathbf{A}^{T}\mathbf{W}_{2}\left[\mathbf{C}_{3}%
\mathbf{\mathbf{W}}_{1}\mathbf{\eta}_{1}-\left[  \mathbf{I}-\mathbf{C}_{3}\mathbf{W}_{2}\right] \mathbf{g}\right] \\
\mathbf{W}=\mathbf{A}^{T}\mathbf{W}_{2}\left[  \mathbf{I}-\mathbf{C}%
_{3}\mathbf{W}_{2}\right]  \mathbf{A}, \mathbf{C}_{3}\triangleq\left[\mathbf{W}_{1}+\mathbf{W}_{2}\right]  ^{-1}%
\end{array}
$\\\hline
\end{tabular}
\caption{Mathematical rules for the evaluation of the message $m_{out}(\mathbf{x}_{2})$, emerging from a function node $f(\mathbf{x}_{1}, \mathbf{x}_{2})$ on the basis of the input message $m_{in,1}(\mathbf{x}_{1})$; note that
in formula no. 4 $N$ denotes the size of the vector $\mathbf{x}_{1}$, and that both  $m_{out}(\mathbf{x}_{2})$ and $f(\mathbf{x}_{1}, \mathbf{x}_{2})$
are independent of $\mathbf{x}_{2}$.
\label{Table2}}%
\end{table*}

1. \emph{Time update} for $\mathbf{x}_{l}^{(L)}$ - Compute the message (see
T2-5, (\ref{m_fp_N}) and (\ref{m_be_L}))%
\begin{align}
\overset{\leftarrow}{m}_{1,j}\left(  \mathbf{x}_{l}^{(L)}\right)   &
=\int\int\,f\left(  \mathbf{x}_{l+1}^{(L)}\left\vert \mathbf{x}_{l}%
^{(L)},\mathbf{x}_{l}^{(N)}\right.  \right) \nonumber\\
&  \cdot\overset{\leftarrow}{m}_{be}\left(  \mathbf{x}_{l+1}^{(L)}\right)
\vec{m}_{fp,j}\left(  \mathbf{x}_{l}^{(N)}\right)  d\mathbf{x}_{l+1}%
^{(L)}d\mathbf{x}_{l}^{(N)}\nonumber\\
&  =\mathcal{N}\left(\mathbf{x}_{l}^{(L)};\mathbf{\eta}_{1,l,j}%
^{(L)},\mathbf{C}_{1,l,j}^{(L)}\right)  ,\label{m_1_L}%
\end{align}
where%
\begin{align}
\mathbf{w}_{1,l,j}^{(L)}  & \triangleq\mathbf{W}_{1,l,j}^{(L)}\mathbf{\eta
}_{1,l,j}^{(L)}=\left(  \mathbf{A}_{l,j}^{(L)}\right)  ^{T}\mathbf{W}%
_{w}^{(L)}\nonumber\\
& \cdot\left[  \mathbf{\bar{C}}_{l+1}\mathbf{w}_{be,l+1}^{(L)}-\mathbf{P}%
_{l}^{(L)}\mathbf{f}_{l,j}^{(L)}\right]  ,\label{w_1_L}%
\end{align}%
\begin{equation}
\mathbf{W}_{1,l,j}^{(L)}\triangleq\left(  \mathbf{C}_{1,l,j}^{(L)}\right)
^{-1}=\left(  \mathbf{A}_{l,j}^{(L)}\right)  ^{T}\mathbf{W}_{w}^{(L)}%
\mathbf{P}_{l}^{(L)}\mathbf{A}_{l,j}^{(L)},\label{W_1_L}%
\end{equation}
$\mathbf{A}_{l,j}^{(L)}\triangleq\mathbf{A}_{l}^{(L)}(\mathbf{x}%
_{l/(l-1),j}^{(N)})$, $\mathbf{W}_{w}^{(L)}\triangleq(\mathbf{C}_{w}%
^{(L)})^{-1}$, $\mathbf{P}_{l}^{(L)}\triangleq\mathbf{I}_{D_{L}}%
-\mathbf{\bar{C}}_{l+1}\mathbf{W}_{w}^{(L)}$, $\mathbf{\bar{C}}_{l+1}%
\triangleq(\mathbf{W}_{w}^{(L)}+\mathbf{W}_{be,l+1}^{(L)})^{-1}$,
$\mathbf{W}_{be,l+1}^{(L)}\triangleq(\mathbf{C}_{be,l+1}^{(L)})^{-1}$,
$\mathbf{f}_{l,j}^{(L)}\triangleq\mathbf{f}_{l}^{(L)}(\mathbf{x}%
_{l/(l-1),j}^{(N)})$ and $\mathbf{w}_{be,l+1}^{(L)}\triangleq\mathbf{W}%
_{be,l+1}^{(L)}\mathbf{\eta}_{be,l+1}^{(L)}$.

2. \emph{Measurement update} for $\mathbf{x}_{l}^{(L)}$ - Compute: a) the
message
\begin{equation}
\vec{m}_{j}\left(  \mathbf{z}_{l}^{(L)}\right)  =f\left(  \mathbf{z}_{l}%
^{(L)}\left\vert \mathbf{x}_{l/(l-1),j}^{(N)},\mathbf{\tilde{x}}_{l+1}%
^{(N)}\right.  \right)  =\delta\left(  \mathbf{z}_{l}^{(L)}-\mathbf{z}%
_{l,j}^{(L)}\right)  ,\label{m_z_L_n}%
\end{equation}
where $\mathbf{z}_{l,j}^{(L)}\triangleq\mathbf{x}_{be,l+1}^{(N)}%
-\mathbf{f}_{l,j}^{(N)}$ and $\mathbf{f}_{l,j}^{(N)}\triangleq\mathbf{f}%
_{l}^{(N)}(\mathbf{x}_{l/(l-1),j}^{(N)})$; b) the messages (see T2-3, T1-3,
T2-2 and T1-2, respectively; see also (\ref{m_fp_N}), (\ref{m_1_L}) and (\ref{m_z_L_n}))%
\begin{align}
\overset{\leftarrow}{m}_{2,j}\left(  \mathbf{x}_{l}^{(L)}\right)   &
=\int\int\,f\left(  \mathbf{z}_{l}^{(L)}\left\vert \mathbf{x}_{l}%
^{(L)},\mathbf{x}_{l}^{(N)}\right.  \right) \nonumber\\
&  \cdot\overset{\leftarrow}{m}_{j}\left(  \mathbf{z}_{l}^{(L)}\right)
\,\vec{m}_{fp,j}\left(  \mathbf{x}_{l}^{(N)}\right)  d\mathbf{x}_{l}%
^{(N)}d\mathbf{z}_{l}^{(L)}\nonumber\\
&  =\mathcal{\mathcal{N}}\left(  \mathbf{z}_{l,j}^{(L)};\mathbf{A}_{l,j}%
^{(N)}\,\mathbf{x}_{l}^{(L)},\mathbf{C}_{w}^{(N)}\right)  ,\label{m_2_Lb}%
\end{align}%
\begin{align}
\overset{\leftarrow}{m}_{3,j}\left(  \mathbf{x}_{l}^{(L)}\right)   &
=\overset{\leftarrow}{m}_{1,j}\left(  \mathbf{x}_{l}^{(L)}\right)
\,\overset{\leftarrow}{m}_{2,j}\left(  \mathbf{x}_{l}^{(L)}\right) \nonumber\\
&  =\mathcal{N}\left(  \mathbf{x}_{l}^{(L)};\mathbf{\eta}_{3,l,j}%
^{(L)},\mathbf{C}_{3,l,j}^{(L)}\right)  ,\label{m_3_L}%
\end{align}%
\begin{align}
\overset{\leftarrow}{m}_{4,j}\left(  \mathbf{x}_{l}^{(L)}\right)   &  =\int
f\left(  \mathbf{y}_{l}\left\vert \mathbf{x}_{l}^{(N)},\,\mathbf{x}_{l}%
^{(L)}\right.  \right)  \,\vec{m}_{fp,j}\left(  \mathbf{x}_{l}^{(N)}\right)
d\mathbf{x}_{l}^{(N)}\nonumber\\
&  =\mathcal{N}\left(  \mathbf{y}_{l};\mathbf{B}_{l,j}\,\mathbf{x}_{l}%
^{(L)}+\mathbf{h}_{l,j},\mathbf{C}_{e}\right) \label{m_4_L_b}\\
&  \equiv\mathcal{N}\left(  \mathbf{x}_{l}^{(L)};\mathbf{\eta}_{4,l,j}%
^{(L)},\mathbf{C}_{4,l,j}^{(L)}\right) \label{m_4_L_a}%
\end{align}
and%
\begin{align}
\overset{\leftarrow}{m}_{be,j}\left(  \mathbf{x}_{l}^{(L)}\right)   &
=\overset{\leftarrow}{m}_{3,j}\left(  \mathbf{x}_{l}^{(L)}\right)
\overset{\leftarrow}{m}_{4,j}\left(  \mathbf{x}_{l}^{(L)}\right) \nonumber\\
&  =\mathcal{N}\left(  \mathbf{x}_{l}^{(L)};\mathbf{\eta}_{be,l,j}%
^{(L)},\mathbf{C}_{be,l,j}^{(L)}\right)  .\label{m_5_L}%
\end{align}
Here,%
\begin{equation}
\mathbf{w}_{3,l,j}^{(L)}\triangleq\mathbf{W}_{3,l,j}^{(L)}\mathbf{\eta
}_{3,l,j}^{(L)}=\mathbf{w}_{1,l,j}^{(L)}+\left(  \mathbf{A}_{l,j}%
^{(N)}\right)  ^{T}\mathbf{W}_{w}^{(N)}\mathbf{z}_{l,j}^{(L)},\label{w_3_L}%
\end{equation}%
\begin{equation}
\mathbf{W}_{3,l,j}^{(L)}\triangleq\left(  \mathbf{C}_{3,l,j}^{(L)}\right)
^{-1}=\mathbf{W}_{1,l,j}^{(L)}+\left(  \mathbf{A}_{l,j}^{(N)}\right)
^{T}\mathbf{W}_{w}^{(N)}\mathbf{A}_{l,j}^{(N)}\text{,}\label{W_3_L}%
\end{equation}
$\mathbf{A}_{l,j}^{(N)}\triangleq\mathbf{A}_{l}^{(N)}(\mathbf{x}%
_{l/(l-1),j}^{(N)})$, $\mathbf{W}_{w}^{(N)}\triangleq\lbrack\mathbf{C}%
_{w}^{(N)}]^{-1}$,
\begin{equation}
\mathbf{w}_{4,l,j}^{(L)}\triangleq\mathbf{W}_{4,l,j}^{(L)}\mathbf{\eta
}_{4,l,j}^{(L)}=\left(  \mathbf{B}_{l,j}\right)  ^{T}\mathbf{W}_{e}\left(
\mathbf{y}_{l}-\mathbf{h}_{l,j}\right)  ,\label{w_4_L}%
\end{equation}%
\begin{equation}
\mathbf{W}_{4,l,j}^{(L)}\triangleq\left(  \mathbf{C}_{4,l,j}^{(L)}\right)
^{-1}=\left(  \mathbf{B}_{l,j}\right)  ^{T}\mathbf{W}_{e}\mathbf{B}%
_{l,j},\label{W_4_L}%
\end{equation}
$\mathbf{B}_{l,j}\triangleq\mathbf{B}_{l}(\mathbf{x}_{l/(l-1),j}^{(N)})$,
$\mathbf{h}_{l,j}\triangleq\mathbf{h}_{l}(\mathbf{x}_{l/(l-1),j}^{(N)})$,
$\mathbf{W}_{e}\triangleq\mathbf{C}_{e}^{-1}$,%
\begin{equation}
\mathbf{w}_{be,l,j}^{(L)}\triangleq\mathbf{W}_{be,l,j}^{(L)}\mathbf{\eta
}_{be,l,j}^{(L)}=\mathbf{w}_{3,l,j}^{(L)}+\mathbf{w}_{4,l,j}^{(L)}%
\label{w_5_L}%
\end{equation}
and%
\begin{equation}
\mathbf{W}_{be,l,j}^{(L)}\triangleq\left(  \mathbf{C}_{be,l,j}^{(L)}\right)
^{-1}=\mathbf{W}_{3,l,j}^{(L)}+\mathbf{W}_{4,l,j}^{(L)}.\label{W_5_L}%
\end{equation}

3. \emph{Merge of forward and backward messages} about $\mathbf{x}_{l}^{(L)}$ -
Compute the message (see (\ref{eq:smooth_dens_1}), (\ref{m_fp_L}),
(\ref{m_5_L}), T1-2 and Fig. \ref{Fig_4})
\begin{align}
m_{sm,j}\left(  \mathbf{x}_{l}^{(L)}\right)   &  =\vec{m}_{fp,j}\left(
\mathbf{x}_{l}^{(L)}\right)  \overset{\leftarrow}{m}_{be,j}\left(
\mathbf{x}_{l}^{(L)}\right) \nonumber\\
&  =\mathcal{N}\left(  \mathbf{x}_{l}^{(L)};\mathbf{\eta}_{sm,l,j}%
^{(L)},\mathbf{C}_{sm,l,j}^{(L)}\right)  \text{,}\label{m_fb_L}%
\end{align}
where%
\begin{equation}
\mathbf{W}_{sm,l,j}^{(L)}\triangleq\left(  \mathbf{C}_{sm,l,j}^{(L)}\right)
^{-1}=\mathbf{W}_{fp,l,j}^{(L)}+\mathbf{W}_{be,l,j}^{(L)},\label{W_fb_L}%
\end{equation}%
\begin{equation}
\mathbf{w}_{sm,l,j}^{(L)}\triangleq\mathbf{W}_{sm,l,j}^{(L)}\mathbf{\eta
}_{sm,l,j}^{(L)}=\mathbf{w}_{fp,l,j}^{(L)}+\mathbf{w}_{be,l,j}^{(L)}%
\text{,}\label{w_fb_L}%
\end{equation}
$\mathbf{W}_{fp,l,j}^{(L)}\triangleq(\mathbf{C}_{fp,l,j}^{(L)})^{-1}$ and
$\mathbf{w}_{fp,l,j}^{(L)}\triangleq\mathbf{W}_{fp,l,j}^{(L)}\mathbf{\eta
}_{fp,l,j}^{(L)}$.

4. \emph{Time update} for $\mathbf{x}_{l}^{(N)}$ - Compute the message (see
T2-1, (\ref{m_be_N}) and (\ref{m_fb_L}))%
\begin{align}
\overset{\leftarrow}{m}_{1,j}\left(  \mathbf{x}_{l}^{(N)}\right)  \, &
=\int\int\,f\left(  \mathbf{x}_{l+1}^{(N)}\left\vert \mathbf{x}_{l}%
^{(L)},\mathbf{x}_{l/(l-1),j}^{(N)}\right.  \right) \nonumber\\
&  \cdot\overset{\leftarrow}{m}_{be}\left(  \mathbf{x}_{l+1}^{(N)}\right)
\,m_{sm,j}\left(  \mathbf{x}_{l}^{(L)}\right)  d\mathbf{x}_{l}^{(L)}%
d\mathbf{x}_{l+1}^{(N)}\nonumber\\
&  =\mathcal{N}\left(  \mathbf{x}_{be,l+1}^{(N)};\mathbf{\eta}_{1,l,j}%
^{(N)},\mathbf{C}_{1,l,j}^{(N)}\right)  \triangleq w_{1,l,j}%
,\label{m_bp_x_N_l+1_exact}%
\end{align}
where%
\begin{equation}
\mathbf{\eta}_{1,l,j}^{(N)}=\mathbf{A}_{l,j}^{(N)}\mathbf{\eta}_{sm,l,j}%
^{(N)}+\mathbf{f}_{l,j}^{(N)}\label{eta_bp}%
\end{equation}
and%
\begin{equation}
\mathbf{C}_{1,l,j}^{(N)}\triangleq\mathbf{A}_{l,j}^{(N)}\mathbf{C}%
_{sm,l,j}^{(N)}\left(  \mathbf{A}_{l,j}^{(N)}\right)  ^{T}+\mathbf{C}%
_{w}^{(N)}.\label{cov_bp}%
\end{equation}

5. \emph{Measurement update} for $\mathbf{x}_{l}^{(N)}$ - Compute: a) the
message
\begin{equation}
\overset{\leftarrow}{m}_{j}\left(  \mathbf{z}_{l}^{(N)}\right)  =\mathcal{N}%
\left(  \mathbf{x}_{l}^{(N)};\mathbf{\eta}_{\mathbf{z},l,j}^{(N)}%
,\mathbf{C}_{\mathbf{z},l,j}^{(N)}\right) \label{m_z_N}%
\end{equation}
and the message (see T3-1)
\begin{align}
\overset{\leftarrow}{m}_{2,j}\left(  \mathbf{x}_{l}^{(N)}\right)   &
=\int\overset{\leftarrow}{m}_{j}\left(  \mathbf{z}_{l}^{(N)}\right)  f\left(
\mathbf{z}_{l}^{(N)}\left\vert \mathbf{x}_{l/(l-1),j}^{(N)}\right.  \right)
d\mathbf{z}_{l}^{(N)}\nonumber\\
&  =K_{l,j}\exp\left[  \frac{1}{2}\left(  \left(  \mathbf{\eta}_{2,l,j}%
^{(N)}\right)  ^{T}\mathbf{W}_{2,l,j}^{(N)}\mathbf{\eta}_{2,l,j}^{(N)}\right.
\right. \nonumber\\
&  -\left.  \left.  \left(  \mathbf{\eta}_{\mathbf{z},l,j}^{(N)}\right)
^{T}\mathbf{W}_{\mathbf{z},l,j}^{(N)}\mathbf{\eta}_{\mathbf{z},l,j}%
^{(N)}-\left(  \mathbf{f}_{l,j}^{(L)}\right)  ^{T}\mathbf{W}_{w}%
^{(L)}\mathbf{f}_{l,j}^{(L)}\right)  \right] \nonumber\\ \triangleq\,w_{2,l,j},\label{m_2_N}%
\end{align}
where%
\begin{equation}
\mathbf{\eta}_{\mathbf{z},l,j}^{(N)}\triangleq\mathbf{\eta}_{be,l+1}%
^{(L)}-\mathbf{A}_{l,j}^{(L)}\mathbf{\eta}_{sm,l,j}^{(L)},\label{eta_z_N}%
\end{equation}%
\begin{equation}
\mathbf{C}_{\mathbf{z},l,j}^{(N)}\triangleq\mathbf{C}_{be,l+1}^{(L)}%
-\mathbf{A}_{l,j}^{(L)}\mathbf{C}_{sm,l,j}^{(L)}\left(  \mathbf{A}_{l,j}%
^{(L)}\right)  ^{T},\label{C_z_N}%
\end{equation}%
\begin{equation}
\mathbf{W}_{2,l,j}^{(N)}\triangleq\left(  \mathbf{C}_{2,l,j}^{(N)}\right)
^{-1}=\mathbf{W}_{\mathbf{z},l,j}^{(N)}+\mathbf{W}_{w}^{(L)},\label{W_2_l_N}%
\end{equation}%
\begin{equation}
\mathbf{w}_{2,l,j}^{(N)}\triangleq\mathbf{W}_{2,l,j}^{(N)}\mathbf{\eta
}_{2,l,j}^{(N)}=\mathbf{w}_{\mathbf{z},l,j}^{(N)}+\mathbf{W}_{w}%
^{(L)}\mathbf{f}_{l,j}^{(L)}\text{,}\label{w_2_l_N}%
\end{equation}
$K_{l,j}=(\det(\mathbf{C}_{\mathbf{z},l,j}^{(N)}+\mathbf{C}_{w}^{(L)}%
))^{-D_{L}/2}$, $\mathbf{W}_{\mathbf{z},l,j}^{(N)}\triangleq(\mathbf{C}%
_{\mathbf{z},l,j}^{(N)})^{-1}$ and $\mathbf{w}_{\mathbf{z},l,j}^{(N)}%
\triangleq\mathbf{W}_{\mathbf{z},l,j}^{(N)}\mathbf{\eta}_{\mathbf{z}%
,l,j}^{(N)}$; b) the messages (see T1-1 and T2-1, respectively)%
\begin{align}
\overset{\leftarrow}{m}_{3,j}\left(  \mathbf{x}_{l}^{(N)}\right)   &
=\overset{\leftarrow}{m}_{1,j}\left(  \mathbf{x}_{l}^{(N)}\right)
\overset{\leftarrow}{m}_{2,j}\left(  \mathbf{x}_{l}^{(N)}\right)  =\nonumber\\
&  =w_{1,l,j}\cdot w_{2,l,j}\triangleq w_{3,l,j}\label{m_3_N}%
\end{align}
and%
\begin{align}
\overset{\leftarrow}{m}_{4,j}\left(  \mathbf{x}_{l}^{(N)}\right)   &  =\int
f\left(  \mathbf{y}_{l}\left\vert \mathbf{x}_{l/(l-1),j}^{(N)},\,\mathbf{x}%
_{l}^{(L)}\right.  \right)  m_{sm,j}\left(  \mathbf{x}_{l}^{(L)}\right)
d\mathbf{x}_{l}^{(L)}\nonumber\\
&  =\mathcal{N}\left(  \mathbf{y}_{l};\mathbf{\eta}_{4,l,j}^{(N)}%
,\mathbf{C}_{4,l,j}^{(N)}\right)  \triangleq w_{4,l,j},\label{m_4_N}%
\end{align}
where $\mathbf{\eta}_{4,l,j}^{(N)}=\mathbf{B}_{l,j}\mathbf{\eta}_{sm,l,j}^{(L)}+\mathbf{h}_{l,j}$ and $\mathbf{C}_{4,l,j}^{(N)}=\mathbf{B}_{l,j}\mathbf{C}_{sm,l,j}^{(L)}(\mathbf{B}_{l,j})^{T}+\mathbf{C}_{e}$.

6. \emph{Merge of forward and backward messages} about $\mathbf{x}%
_{l}^{(N)}$ - This requires: a) computing the messages (see (\ref{m_3_N}) and
(\ref{m_4_N}))
\begin{align}
\overset{\leftarrow}{m}_{be,j}\left(  \mathbf{x}_{l}^{(N)}\right)   &
=\overset{\leftarrow}{m}_{3,j}\left(  \mathbf{x}_{l}^{(N)}\right)
\overset{\leftarrow}{m}_{4,j}\left(  \mathbf{x}_{l}^{(N)}\right)
\,\nonumber\\
&  =w_{3,l,j}\cdot w_{4,l,j}=w_{1,l,j}\cdot w_{2,l,j}\cdot w_{4,l,j}\triangleq
W_{l,j}\label{W_final}%
\end{align}
and (see (\ref{m_fp_N}) and T1-1)
\begin{align}
\overset{\leftarrow}{m}_{sm,j}\left(  \mathbf{x}_{l}^{(N)}\right)   &
=\,\vec{m}_{fp,j}\left(  \mathbf{x}_{l}^{(N)}\right)  \overset{\leftarrow}%
{m}_{be,j}\left(  \mathbf{x}_{l}^{(N)}\right) \nonumber\\
&  =W_{l,j}\,\delta\left(  \mathbf{x}_{l}^{(N)}-\mathbf{x}_{l/(l-1),j}%
^{(N)}\right)  ;\label{m_5_Nbis}%
\end{align}
b) normalising the weight set $\{W_{l,j}\,\}$, i.e. generating the new weight%
\begin{equation}
W_{sm,l,j}\,\triangleq W_{l,j}/\sum_{j=0}^{N_{p}-1}W_{l,j}\label{W_norm}%
\end{equation}
for $j=0,1,,...,N_{p}-1$; c) setting
\begin{equation}
m_{sm,j}\left(  \mathbf{x}_{l}^{(N)}\right)  =W_{sm,l,j}\,\delta\left(
\mathbf{x}_{l}^{(N)}-\mathbf{x}_{l/(l-1),j}^{(N)}\right)  \,\label{m_sm_N}%
\end{equation}
for $j=0,1,,...,N_{p}-1$.

7. \emph{Generation of smoothed pdfs and input messages for the next recursion
}- Compute: a) the pdfs%
\begin{align}
\hat{f}\left(\mathbf{x}_{l},|\mathbf{y}_{1:N}\right)  & \triangleq%
\sum\limits_{j=0}^{N_{p}-1}
\,\,m_{sm,j}\left(\mathbf{x}_{l}^{(N)}\right)  \,m_{sm,j}\left(
\mathbf{x}_{l}^{(L)}\right)\label{smmothed_pdf_LN}%
\end{align}%
\begin{equation}
\hat{f}\left(\mathbf{x}_{l}^{(N)}|\mathbf{y}_{1:N}\right) \triangleq
\sum\limits_{j=0}^{N_{p}-1}
\,\,m_{sm,j}\left(\mathbf{x}_{l}^{(N)}\right) \label{smoothed_pfd_N}%
\end{equation}
and%
\begin{equation}
\hat{f}\left(\mathbf{x}_{l}^{(L)}|\mathbf{y}_{1:N}\right) \triangleq
\sum\limits_{j=0}^{N_{p}-1}
W_{sm,l,j}\,   m_{sm,j}\left(
\mathbf{x}_{l}^{(L)}\right),\label{smoothed_pfd_L}%
\end{equation}
that represent approximations of the marginal smoothed pdfs of $\mathbf{x}_{l},
$ $\mathbf{x}_{l}^{(N)}$ and $\mathbf{x}_{l}^{(L)}$, respectively; b) the
input messages
\begin{equation}
\overset{\leftarrow}{m}_{be}\left(  \mathbf{x}_{l}^{(N)}\right)
\triangleq\delta\left(\mathbf{x}_{l}^{(N)}-\mathbf{x}_{be,l}^{(N)}\right)
\label{m_be_N_l}%
\end{equation}
and
\begin{equation}
\overset{\leftarrow}{m}_{be}\left(\mathbf{x}_{l}^{(L)}\right)
=\mathcal{N}\left(\mathbf{x}_{l}^{(L)};\mathbf{\eta}_{be,l}^{(L)}%
,\mathbf{C}_{be,l}^{(L)}\right) \label{m_be_L_l}%
\end{equation}
for the next recursion; here,%
\begin{equation}
\mathbf{x}_{be,l}^{(N)}\triangleq%
\sum\limits_{j=0}^{N_{p}-1}
W_{sm,l,j}\,\mathbf{x}_{be,l,j}^{(N)},\label{ext_x_N}%
\end{equation}%
\begin{equation}
\mathbf{\eta}_{be,l}^{(L)}\triangleq\sum\limits_{j=0}^{N_{p}-1} W_{sm,l,j}\,\mathbf{\eta}_{sm,l,j}^{(L)}
\label{eta_cond}
\end{equation}%
and%
\begin{align}
\mathbf{C}_{be,l}^{(L)} &  \triangleq%
\sum\limits_{j=0}^{N_{p}-1}
W_{sm,l,j}\,\mathbf{C}_{sm,l,j}^{(L)}\nonumber\\
&  +%
{\displaystyle\sum\limits_{j=0}^{N_{p}-1}}
W_{sm,l,j}\left(  \,\mathbf{\eta}_{sm,l,j}^{(L)}-\mathbf{\eta}_{be,l}%
^{(L)}\right)  \left(  \,\mathbf{\eta}_{sm,l,j}^{(L)}-\mathbf{\eta}%
_{be,l}^{(L)}\right)  ^{T}\label{cov_cond}%
\end{align}

After completing step 7, the $(T-l)$-th recursion of the RBSS technique is
over. Then, the recursion index $l$ is decreased by one; if it equals zero,
the backward pass is over, otherwise a new recursion is started. Note also
that the first recursion of the backward pass requires the knowledge of its
input messages $\overset{\leftarrow}{m}_{be}(\mathbf{x}_{T}^{(N)})$ and
$\overset{\leftarrow}{m}_{be}(\mathbf{x}_{T}^{(L)})$, whose evaluation is
based on the statistical information generated in the last recursion of the
forward pass. In fact, in our work these messages are defined as
\begin{equation}
\overset{\leftarrow}{m}_{be}\left(\mathbf{x}_{T}^{(N)}\right)
\triangleq\delta\left(\mathbf{x}_{T}^{(N)}-\mathbf{x}_{fe,T}^{(N)}\right)
\label{init_N}%
\end{equation}
and
\begin{equation}
\overset{\leftarrow}{m}_{be}\left(\mathbf{x}_{T}^{(L)}\right)
\triangleq\mathcal{N}\left(\mathbf{x}_{T}^{(L)};\mathbf{\eta}_{fe,T}%
^{(L)},\mathbf{C}_{fe,T}^{(L)}\right)  ,\label{init_L}%
\end{equation}
respectively; here, $\mathbf{x}_{fe,T}^{(N)}\triangleq\sum\limits_{j=0}^{N_{p}-1} w_{T/T,j}\,\mathbf{x}_{T/(T-1),j}^{(N)}$, whereas the parameters $\mathbf{\eta}_{fe,T}^{(L)}$ and $\mathbf{C}%
_{fe,T}^{(L)}$ of (\ref{init_L}) are evaluated on the basis of formulas
(\ref{eta_cond}) and (\ref{cov_cond}), but employing, in place of the Gaussian
messages $\{\overset{\leftarrow}{m}_{be,j}(\mathbf{x}_{l}^{(N)})\}$ (see
(\ref{W_final})), the messages $\{\mathcal{N}(\mathbf{x}_{T}^{(L)}%
;\mathbf{\eta}_{fe,T,j}^{(L)},\mathbf{C}_{fe,T,j}^{(L)})\}$ generated by the
MU for the linear state component in the last (i.e., in the $T$-th) recursion
of FF.

The RBSS\ algorithm illustrated above deserves various comments, that are
listed below.

\begin{enumerate}
\item The message flow in the backward pass proceeds in a \emph{reverse order}
with respect to the forward pass (a similar scheduling in
the backward pass has been adopted in \cite{Fong_2002}); in fact, in MPF the
evaluation of particle weights and the prediction of new particles for the
next recursion (accomplished in the MU\ and in the TU, respectively, for the
nonlinear state component) precedes the MU and the TU\ for the linear state
component. Moreover, unlike TF, a \emph{single pass} is accomplished over the
FG.

\item In step 1 a \emph{one-step ahead prediction} is evaluated for
$\mathbf{x}_{l}^{(L)}$ on the basis of the pdf of $\mathbf{x}_{l+1}^{(L)}$
(provided by the particle-independent message $\overset{\leftarrow}{m}%
_{be}(\mathbf{x}_{l+1}^{(L)})$ (\ref{m_be_L})). A conceptually similar task is
carried out for $\mathbf{x}_{l}^{(N)}$ in step 4. However, in the last case,
pdf prediction does not involve the generation of new particles (like in the
TU step of MPF), but only the computation of \emph{new weights} for the
particles originating from the forward pass. For this reason, the
\emph{support} of the pdf $\hat{f}(\mathbf{x}_{l}^{(N)}|\mathbf{y}_{1:N})$
(\ref{smoothed_pfd_N}) estimated for $\mathbf{x}_{l}^{(N)}$ in the backward
pass remains exactly the same as that of the corresponding filtered pdf
computed in the forward pass.

\item In step 2 the pdf $\overset{\leftarrow}{m}_{1,j}(\mathbf{x}_{l}^{(L)}) $
(\ref{m_1_L}) emerging from step 1 is refined on the basis of a) the
measurement $\mathbf{y}_{l}$ and b) the pseudo-measurement $\mathbf{z}%
_{l,j}^{(L)}$, which depends on the particle index $j$ through $\mathbf{x}%
_{l,j}^{(N)}$ only (since a \emph{single} particle is available for
$\mathbf{x}_{l+1}^{(N)}$). Even if this entails a \emph{loss of diversity} in
the pseudo-measurement set $\{\mathbf{z}_{l,j}^{(L)}\}$ with respect to the
corresponding set generated by MPF in the forward pass, the use of these
quantities in state estimation is still beneficial. Incidentally, we note that
no attention to the exploitation of pseudo-measurements $\mathbf{z}_{l}^{(L)}$
and $\mathbf{z}_{l}^{(N)}$ is paid in the development of the other RBPS
methods available in the literature, even if these quantities are known to
play an important role in state estimation \cite{Schon_2005},
\cite{Vitetta_2016}, \cite{Vitetta_2017}.

\item In step 3 the merge of the forward message $\vec{m}_{fp,j}%
(\mathbf{x}_{l}^{(L)})$ with the backward message $\overset{\leftarrow}%
{m}_{be,j}(\mathbf{x}_{l}^{(L)})$ results in the `smoothed' message
$m_{sm,j}(\mathbf{x}_{l}^{(L)})$ (\ref{m_fb_L}), which is expected to provide
a more refined statistical representation of $\mathbf{x}_{l}^{(L)}$ than
$\vec{m}_{fp,j}(\mathbf{x}_{l}^{(L)})$ or $\overset{\leftarrow}{m}%
_{be,j}(\mathbf{x}_{l}^{(L)})$ alone (under the assumption that $\mathbf{x}%
_{l}^{(N)}=\mathbf{x}_{l/(l-1),j}^{(L)}$) and, consequently, to improve the
accuracy of the particle weights evaluated in steps 4 and 5; note also that
$m_{sm,j}(\mathbf{x}_{1}^{(L)})=\overset{\leftarrow}{m}_{be,j}(\mathbf{x}%
_{1}^{(L)})$ and $m_{sm,j}(\mathbf{x}_{T}^{(L)})=\vec{m}_{fp,j}(\mathbf{x}%
_{T}^{(L)})$ should be assumed, since at the instant $l=1$ ($l=T$) only a backward
estimate (a forward prediction) is available for $\mathbf{x}_{l}^{(L)}$.

\item In step 3 the equivalence between the expressions (\ref{m_4_L_b}) and
(\ref{m_4_L_a}) is motivated by the fact that they differ by a scale factor
and that scale factors can be always neglected in passing Gaussian messages
\cite{Loeliger_2007}.

\item In step 5 the factors $w_{1,l,j}$, $w_{2,l,j}$ and $w_{4,l,j}$ of
the overall weight $W_{l,j}$ (\ref{W_final}) are related to the state
transition $\mathbf{x}_{l+1}^{(N)}\rightarrow\mathbf{x}_{l}^{(N)}$, to the
statistical representation of $\mathbf{z}_{l}^{(N)}$ (conveyed by the Gaussian
message $\overset{\leftarrow}{m}_{j}(\mathbf{z}_{l}^{(N)})$ (\ref{m_z_N})) and
to the measurement $\mathbf{y}_{l}$, respectively. Note also that: a) the weight
$w_{1,l,j}$ depends on the (particle-independent) estimate $\mathbf{x}%
_{be,l+1}^{(L)}$, which can be interpreted as an additional pseudo-measurement
originating from our knowledge of the future (and, consequently, unavailable
in the forward pass); b) the weight $w_{2,l,j}$ (\ref{m_2_N}) cannot be
computed in the forward pass because of the scheduling adopted in
MPF (the TU for the nonlinear state component represents the last step
accomplished in each recursion of MPF); c) the weight $w_{4,l,j}$ corresponds
to the weight $w_{l/l,j}$ computed by MPF in the forward pass but, as already
mentioned at point 4), is expected to be more accurate thanks to the
availability of more refined statistical information about $\mathbf{x}%
_{l}^{(L)}$ (conveyed by the message $m_{sm,j}(\mathbf{x}_{l}^{(L)})$
(\ref{m_fb_L}) in place of $m_{fp,j}(\mathbf{x}_{l}^{(L)})$ (\ref{m_fp_L})).

\item Steps 1-6 need to be repeated $N_{p}$ times, once for each particle of
the set $\{\mathbf{x}_{l/(l-1),j}^{(N)}\}$; in practice, this task can be
\emph{parallelized}, since the processing executed for any particle within
these steps is not influenced from that carried out for all the other particles.

\item The expressions of the weights $w_{1,l,j}$, $w_{2,l,j}$ and $w_{4,l,j}$
have similar mathematical structure (see (\ref{m_bp_x_N_l+1_exact}),
(\ref{m_2_N}) and (\ref{m_4_N}), respectively) in the sense that they are
given by the product of an exponential with a particle-dependent factor. An
approximate evaluation of these weights can be obtained neglecting the
contribution of a such a factor in each of their expressions. As a matter of
fact, our computer simulations have evidenced that, at least for the
considered SSM, this simplification does not entail a visible loss in RBSS
accuracy. However, if used, it requires the adoption of weight
normalization for each of the three weight sets; consequently, the overall
weight $W_{l,j}$ (see (\ref{W_final})) is computed as
\begin{equation}
W_{l,j}\,=\tilde{w}_{1,l,j}\cdot\tilde{w}_{2,l,j}\cdot\tilde{w}_{4,l,j}%
\text{,}\label{W_final_bis}%
\end{equation}
where $\tilde{w}_{k,l,j}\,\triangleq w_{k,l,j}/\sum_{j=0}^{N_{p}-1}w_{k,l,j}$ 
for $k=1,2$ and $4$.

\item The final particle weights $\{W_{sm,l,j}\}$ (see (\ref{W_norm})) are
employed to generate both the final estimate $\mathbf{x}_{be,l}^{(N)}$
(\ref{ext_x_N}) of $\mathbf{x}_{l}^{(N)}$ and the $N_{p}$-component
\emph{Gaussian mixture} (GM) $\hat{f}(\mathbf{x}_{l}^{(L)}|\mathbf{y}_{1:N})$
(\ref{smoothed_pfd_L}), expressing our final estimate of the pdf of
$\mathbf{x}_{l}^{(L)}$. This GM, however, is not passed to the next recursion
as it is, since this would be make the complexity of our message passing
algorithm unmanageable. This is the reason why this pdf is condensed in the
Gaussian message $\overset{\leftarrow}{m}_{be}(\mathbf{x}_{l}^{(L)})$
(\ref{m_be_L_l}) by means of a standard transformation, expressed by formulas
(\ref{eta_cond}) and (\ref{cov_cond}), and preserving both the mean and the
covariance matrix of the GM itself (e.g., see \cite[Sect. 4]{Runnalls_2007}).
\end{enumerate}

Our final comment concerns the smoothing of the linear state component and has
been inspired by the considerations illustrated in \cite[Par. IV-D]%
{Lindsten_2016}, where it is stressed that in Rao-Blackwellized methods the
statistics for the linear state component need to be computed
\emph{conditionally} on the considered nonlinear state trajectories. As a
matter of fact, our RBSS algorithm generates a single estimate of
\emph{nonlinear state trajectory} in its backward pass (the $l$-th point of
this trajectory is represented by $\mathbf{x}_{be,l}^{(N)}$ with
$l=1,2,...,T-1$ and by $\mathbf{x}_{fe,T}^{(N)}$ for $l=T$); however, the
statistical models for the linear state components associated with this
trajectory (see $\hat{f}(\mathbf{x}_{l}^{(L)}|\mathbf{y}_{1:N})$
(\ref{smoothed_pfd_L}) or its condensed representation $\overset{\leftarrow
}{m}_{be}(\mathbf{x}_{l}^{(L)}) $ (\ref{m_be_L_l})) do not satisfy the above
mentioned condition, since they do not actually refer to a specific nonlinear
state trajectory. This suggests that, once the RBSS algorithm has been carried
out, more refined statistics for the linear state component could be computed by:

\begin{enumerate}
\item Carrying out, first of all, a new forward pass under the assumption that
the nonlinear state component is \emph{known} and, in particular,
$\mathbf{x}_{l}^{(N)}=\mathbf{x}_{be,l}^{(N)}$ for $l=1,2,..,T-1$ and
$\mathbf{x}_{T}^{(N)}=\mathbf{x}_{fe,T}^{(N)}$; this produces a \emph{single}
message $\vec{m}_{fp}(\mathbf{x}_{l}^{(L)})\triangleq\mathcal{N}%
(\mathbf{x}_{l}^{(L)};\mathbf{\eta}_{fp,l}^{(L)},\mathbf{C}_{fp,l}^{(L)})$ in
place of the $N_{p}$ messages $\{\vec{m}_{fp,j}(\mathbf{x}_{l}^{(L)})\}$ (see
(\ref{m_fp_L})) for $l=2,..,T$.

\item Then, accomplishing a new backward pass under the same assumption as
the previous point; this generates a \emph{single} Gaussian message
$\overset{\leftarrow}{m}_{be}(\mathbf{x}_{l}^{(L)})\triangleq\mathcal{N}%
(\mathbf{x}_{l}^{(L)};\mathbf{\eta}_{be,l}^{(L)},\mathbf{C}_{be,l}^{(L)})$ in
place of the $N_{p}$ messages $\{\overset{\leftarrow}{m}_{be,j}(\mathbf{x}%
_{l}^{(L)})\}$ (see (\ref{m_5_L})) for $l=T-1,T-2,..,1$ (note that
$\overset{\leftarrow}{m}_{be}(\mathbf{x}_{T}^{(L)})$ is still given by
(\ref{init_L})).

\item Finally, merging $\vec{m}_{fp}(\mathbf{x}_{l}^{(L)})$ and $\overset
{\leftarrow}{m}_{be}(\mathbf{x}_{l}^{(L)})$ in the message $m_{sm}%
(\mathbf{x}_{l}^{(L)})=\mathcal{N}(\mathbf{x}_{l}^{(L)};\mathbf{\eta}%
_{sm,l}^{(L)},\mathbf{C}_{sm,l}^{(L)})$, with $l=2,1,..,T-1$ ($m_{sm}%
(\mathbf{x}_{1}^{(L)})=\overset{\leftarrow}{m}_{be}(\mathbf{x}_{1}^{(L)})$ and
$m_{sm}(\mathbf{x}_{T}^{(L)})=\vec{m}_{fp}(\mathbf{x}_{T}^{(L)})$ are assumed)
on the basis of (\ref{m_fb_L})-(\ref{w_fb_L}), so that a new final estimate
$\mathbf{\eta}_{sm,l}^{(L)}$ is available for $\mathbf{x}_{l}^{(L)}$.
\end{enumerate}

We believe that, even if this procedure is conceptually appealing, the
improvement it may provide in the estimation accuracy for the linear state
component is influenced by a) the \emph{number of modes} of the density of
$\mathbf{x}_{l}^{(L)}$ (since the adopted \emph{unimodal} model for this state
component might provide a poor statistical representation of it) and b) the
presence of \emph{large errors}, at specific instants, in the estimated
nonlinear state trajectory.

\subsection{Comparison of the RBSS algorithm with other RBPS
methods\label{Comparison_RBPS}}

Despite their substantially different structures, the other RBPS methods
available in the technical literature \cite{Briers_2010}, \cite{Fong_2002},
\cite{Lindsten_2016} share the following relevant features: 1) the computation of
an estimate of the \emph{joint} smoothing density $f(\mathbf{x}_{1:T}%
|\mathbf{y}_{1:T})$; 2) the \emph{reuse} of FF particles and weights; 3) the use of
\emph{resampling} in the generation of backward trajectories; 4) the exploitation
of \emph{Kalman techniques} for the linear state component. In the following
we provide some details about these features, so that some important
differences between such techniques and the RBSS algorithm can be easily understood.

The \emph{first feature} refers to the fact that these techniques aim at
generating \emph{realizations} from the complete \emph{joint} smoothing pdf
$f(\mathbf{x}_{1:T}|\mathbf{y}_{1:T})$. Each realization consists of a) a
trajectory (i.e., a set of $T$ particles, one for each observation instant)
for the nonlinear state component and a set of $T$ Gaussian pdfs (one for each
observation instant) \cite{Briers_2010}, \cite{Lindsten_2016} or b) a
trajectory for the entire state \cite{Fong_2002} (in this case a
particle-based representation is adopted for the linear state component too).
This approach provides the following relevant advantage: any marginal
smoothing density (like those we are interested in) can be easily obtained from the joint density by marginalization (i.e., by discarding the particle sets and the associated Gaussian densities that refer
to the instants we are not interested in). This benefit, however, is obtained
at the price of a substantial computational complexity in all cases. In fact,
the algorithms proposed in \cite[p. 443]{Fong_2002} and \cite[p.
357]{Lindsten_2016} require to be re-run $M$ times, if $M$ realizations of
$f(\mathbf{x}_{1:T}|\mathbf{y}_{1:T})$ are needed; luckily, the processing
accomplished in each run reuses all the particles and the weights computed in
the forward pass. On the contrary, a \emph{single backward pass} is
accomplished in the algorithm derived in \cite[p. 75]{Briers_2010}; this
entails, however, the generation of a new set of weighted particles and
Gaussian densities (representing the nonlinear state component and the linear
state component, respectively); moreover, the evaluation of marginal
smoothing densities is computationally intensive, since it requires merging all
the information (particles, weights and Gaussian densities) emerging from both
passes (see \cite[Par. 4.1.2, p. 80]{Briers_2010}).

The \emph{second feature} concerns the fact that the particles and the
associated weights generated in the forward pass are reused in the backward
pass, even if in different ways. More specifically, in the backward pass of
the RPBS techniques of \cite{Briers_2010} and \cite{Lindsten_2016}, particles
are re-weighted; moreover, each new weight is evaluated as the product of the
weight computed in the forward pass\ for the considered particle with a new
weight generated on the basis of backward statistics (see, in particular, step
3)-b)-ii) of Algorithm 1 in \cite[p. 357]{Lindsten_2016} and the \emph{particle
smoothing} task of Algorithm 4 in \cite[p. 443]{Fong_2002}). On the one hand,
the reuse, in the backward pass, of the particles generated in the forward
pass greatly simplifies BIF. On the other hand, it places a strong constraint
on the \emph{support} of each of the pdfs computed for nonlinear state
component; in fact, such a support is restricted to that identified for the
predicted/filtered pdfs in the forward pass. This is the reason why the RBPS
technique developed in \cite{Briers_2010} includes an algorithm for
generating, in the backward pass, new particles, which are independent of
those computed in the forward pass. The price to be paid for this, however, is
represented by the additional computational load due to 1) particle generation
in the backward pass and b) the complexity of the method employed for merging
forward and backward particles (and their associated weights) to compute the
required smoothed densities (see, in particular, \cite[Par. 4.1.2, p.
80]{Briers_2010}).

As far as the \emph{third feature} is concerned, it is worth mentioning that
the use of resampling in \cite{Fong_2002}, \cite{Lindsten_2016} is
substantially different from that of \cite{Briers_2010}. In fact, in the first
case, resampling is applied to the particle set generated in the TU of each
recursion of the forward pass when evaluating a new trajectory in a backward
pass; this is motivated by the fact that the mechanism of particle selection
can benefit from more refined statistical information, since the new weights
generated in the backward pass for the available particle sets are expected to
be more reliable than those computed in the forward pass. On the contrary, in
the second case, resampling is applied to the new particle set generated in
each recursion of the backward pass, exactly like in the forward pass.

Finally, the \emph{fourth feature} concerns the exploitation of Kalman
techniques and, in particular, of Kalman smoothing for the linear state
component in the considered RBPS algorithms. Note, however, that a different
use of these standard tools is made in the considered manuscripts. In fact, on the one hand, in the RBPS\ techniques proposed in \cite[p. 76]{Briers_2010} and
\cite[p. 443]{Fong_2002} smoothing for linear state component is accomplished
within the backward pass and exploits the statistical information about the
linear state component generated by Rao-Blackwellized filtering in the forward
pass. On the other hand, in \cite{Lindsten_2016} the backward pass aims at
generating a trajectory for the \emph{nonlinear state component only}; such a
trajectory is based on a) the information generated in the forward pass about
this component and b) those generated about the linear state component in the
backward pass only. For this reason, in this case, an additional forward pass
for the linear state component only is accomplished, under the assumption that
the nonlinear state trajectory is known, after that the backward pass has been
completed; finally, Kalman smoothing is carried out to merge forward and
backward information, as illustrated at the end of the previous Paragraph.

From the considerations illustrated above, it can be easily inferred that, on
the one hand, the RBSS algorithm shares feature 4) and part of feature 2) with
the other RBPS techniques (in fact, it reuses the FF particles, but not their
weights). On the other hand, the RBSS algorithm does not share features 1) and
3); this makes it much faster, since both resampling and the generation of
multiple trajectories are time consuming tasks. The other significant
differences between the RBSS algorithm and the other methods can be
summarized as follows. The algorithms developed in \cite{Briers_2010} and
\cite{Fong_2002} apply to a mixed linear/nonlinear SSM whose state equation
for the nonlinear component (see (\ref{eq:XL_update})) does contain the
nonlinear term $\mathbf{f}_{l}^{(L)}(\mathbf{x}_{l}^{(N)})$ (see, in
particular, \cite[eq. (50), p. 75]{Briers_2010} and \cite[eq. (10a), p.
441]{Fong_2002}); consequently, the only alternative method applicable to the
SSM expressed by (\ref{eq:XL_update})-(\ref{eq:y_t}) in its complete form is
represented by the technique devised in \cite{Lindsten_2016}. Moreover, as
mentioned in the previous Paragraph, the RBSS\ algorithm, unlike all the other
RBPS\ methods, fully exploits the available pseudo-measurements.

\subsection{A message passing algorithm for estimating the joint smoothing
density\label{Estimating_joint_pdf}}

Even if backward processing in the RBSS algorithm has been explicitly
devised for estimating the marginal smoothing densities $\{f(\mathbf{x}%
_{l}|\mathbf{y}_{1:T})\}$, the message passing procedure each of its recursion
consists of can be easily modified to generate, like the RBPS method proposed
in \cite{Lindsten_2016}, $M$ (equally likely) \emph{nonlinear} state
trajectories providing a point mass approximation of the joint smoothing pdf
$f(\mathbf{x}_{1:T}^{(N)}|\mathbf{y}_{1:T})$ (e.g., see \cite[eq.
9]{Lindsten_2016}). In practice, this requires: a) accomplishing a single
forward pass (MPF) followed by $M$ distinct backward passes; b) modifying part
of the backward processing devised for RBSS. As far as the last point is
concerned, let us focus, like in Paragraph \ \ref{RBSS_derivation}, on the
$(T-l)$-th recursion of the backward pass (with $l=T-1,T-2,...,1$) of the new
particle smoother (called \emph{enhanced} RBSS, ERBSS, in the following). The
modifications made within the considered recursion originate from the fact
that the nonlinear state trajectory $\{\mathbf{x}_{be,l}^{(N)},l=1,2,...,T\}$
constructed in the ERBSS backward pass consists entirely of particles
generated in the forward pass (and not of a linear combination of them, like
in RBSS; see (\ref{ext_x_N})). For this reason, we set $\mathbf{x}%
_{be,l+1}^{(N)}=$ $\mathbf{x}_{(l+1)/l,j_{l+1}}^{(N)}$ and $(\mathbf{\eta
}_{be,l+1}^{(L)},\mathbf{C}_{be,l+1}^{(L)})=(\mathbf{\eta}_{sm,l+1,j_{l+1}%
}^{(L)}$,$\mathbf{C}_{sm,l+1,j_{l+1}}^{(L)})$ in the \emph{input} messages
$\overset{\leftarrow}{m}_{be}(\mathbf{x}_{l+1}^{(N)})$ (\ref{m_be_N}) and
$\overset{\leftarrow}{m}_{be}(\mathbf{x}_{l+1}^{(L)})$ (\ref{m_be_L}),
respectively, if the specific particle $\mathbf{x}_{(l+1)/l,j_{l+1}}^{(N)}$
has been selected within the particle set $\{\mathbf{x}_{(l+1)/l,j}%
^{(N)},j=0,1,...,N_{p}-1\}$ in the previous (i.e., in the $(T-l-1)$-th)
recursion; the other two input messages $\vec{m}_{fp,j}(\mathbf{x}_{l}^{(N)})$
(\ref{m_fp_N}) and $\vec{m}_{fp,j}(\mathbf{x}_{l}^{(L)})$ (\ref{m_fp_L}),
however, remain unchanged. ERBSS backward processing can be organized
according to seven steps, exactly like RBSS. The first six steps coincide with
steps 1-6 of the RBSS algorithm, whereas the remaining one is described below.

7. \emph{Sample} $\mathbf{x}_{l}^{(N)}$ \emph{and \ generate} \emph{input
messages for the next recursion }- This requires: a) drawing a sample (denoted
$\mathbf{x}_{l/(l-1),j_{l}}^{(N)}$) from the particle set $\{\mathbf{x}%
_{l/(l-1),j}^{(N)}\}$, whose elements are characterized by the probabilities
$\{\Pr\{\mathbf{x}_{l/(l-1),j}^{(N)}\}=W_{sm,l,j}\,\}$; b) setting 
$\mathbf{x}_{be,l}^{(N)}=\mathbf{x}_{l/(l-1),j_{l}}^{(N)}$ 
and $(\mathbf{\eta}_{be,l}^{(L)},\mathbf{C}_{be,l}^{(L)})=(\mathbf{\eta
}_{sm,l,,j_{l}}^{(L)}$,$\mathbf{C}_{sm,l,,j_{l}}^{(L)})$, so that the
nonlinear backward trajectory is extended by one step, and the input messages
$\overset{\leftarrow}{m}_{be}(\mathbf{x}_{l}^{(N)})$ (\ref{m_be_N}) and
$\overset{\leftarrow}{m}_{be}(\mathbf{x}_{l}^{(L)})$ (\ref{m_be_L}) are ready
for the next recursion.

The initialization of the ERBSS algorithm requires the knowledge of its input
messages $\overset{\leftarrow}{m}_{be}(\mathbf{x}_{T}^{(N)})$ and
$\overset{\leftarrow}{m}_{be}(\mathbf{x}_{T}^{(L)})$, that are defined as
\begin{equation}
\overset{\leftarrow}{m}_{be}\left(\mathbf{x}_{T}^{(N)}\right)
\triangleq\delta\left(\mathbf{x}_{T}^{(N)}-\mathbf{x}_{T/(T-1),j_{T}}%
^{(N)}\right) \label{new_init_1}%
\end{equation}
and
\begin{equation}
\overset{\leftarrow}{m}_{be}\left(\mathbf{x}_{T}^{(L)}\right)
\triangleq\mathcal{N}\left(\mathbf{x}_{T}^{(L)};\mathbf{\eta}_{fe,T,j_{T}%
}^{(L)},\mathbf{C}_{fe,T,j_{T}}^{(L)}\right),\label{new_init_2}%
\end{equation}
respectively; here, $\mathbf{x}_{T/(T-1),j_{T}}^{(N)}$ denotes the particle
selected by sampling the particle set $\{\mathbf{x}_{T/(T-1),j}^{(N)}\}$; the
probabilities of its particles are proportional to their weights
$\{w_{T/T,j}\}$ generated by the MPF MU for the nonlinear state
component in its final recursion.

As already mentioned above, the backward pass described above has to be
repeated $M$ times, once for each of the $M$ nonlinear state trajectories; then, smoothing of the linear state component is accomplished for each of them. For this reason, as already explained at the end of Paragraph \ref{RBSS_derivation}, the following tasks are carried out for each nonlinear state trajectory: a)  a new forward pass, followed by a new backward pass, is run for the linear state
component only (under the assumption that the nonlinear state component is
\emph{known}); b) forward prediction and backward estimation messages are merged.

It is worth stressing that the structure of the proposed ERBSS technique is
very similar to that of the Algorithm 2 described in \cite[p. 359]{Lindsten_2016}; the main differences between these two algorithms can be summarized as follows:

\begin{enumerate}
\item The backward processing developed in \cite[p. 359]{Lindsten_2016}
exploits the knowledge of the particle sets/weights
generated in the forward pass, but ignores the associated Gaussian models that represent
 the forward predictions for the linear state
component (actually, the use of such models is limited to the initialization
of the backward simulator). Consequently, step 3 of our RBSS\ algorithm is not
accomplished or, equivalently, (\ref{W_fb_L}) and (\ref{w_fb_L}) are replaced
by $\mathbf{W}_{sm,l,j}^{(L)}\triangleq\mathbf{W}_{be,l,j}^{(L)}$ and
$\mathbf{w}_{sm,l,j}^{(L)}\triangleq\mathbf{w}_{be,l,j}^{(L)}$ , respectively.
From a conceptual viewpoint, two specific motivations can be provided for this
specific choice. The first is represented by the fact that, generally
speaking, the message $\vec{m}_{fp,j}(\mathbf{x}_{l}^{(L)})$ and the message
$\overset{\leftarrow}{m}_{be,j}(\mathbf{x}_{l}^{(L)})$ (see (\ref{m_fp_L}) and
(\ref{m_5_L}), respectively) refer to a specific forward nonlinear trajectory
and to a (unique) backward nonlinear trajectory, respectively, that \emph{do
not merge at the considered instant }(i.e., at the instant $t=l$);
consequently, fusing these densities may result in poor statistical information
and, in particular, may lead to the evaluation of inaccurate weights for the
particle set $\{\mathbf{x}_{l/(l-1),j}^{(N)}\}$. The second motivation is
represented by the fact that statistical (Gaussian) models generated by
backward processing for the linear state component are really
\emph{conditioned} on the selected nonlinear state trajectory; for this
reason, once backward processing is over, a new forward pass \emph{only} has to be
carried for each of the $M$ nonlinear trajectories (in other words, unlike the
ERBSS technique, an additional backward pass is no more required).

\item The particle weights evaluated by Algorithm 2 of \cite[p. 359]%
{Lindsten_2016} in its backward pass are partly based on the weights
$\{w_{l/l,j}\}$ (computed in the forward pass). In particular, the weight
$w_{l/l,j}$ replaces $w_{4,l,j}$ in the expression of the overall weights (see
$W_{l,j}$ (\ref{W_final}) and \cite[Algorithm 1, step 3)-b)-ii), p.
357]{Lindsten_2016}) for any $j$ and $l$.
\end{enumerate}

Actually, our computer simulations have evidenced that particle smoothing
benefits from merging forward and backward information about the linear state
component; in fact, this improves both numerical stability of BIF and its
estimation accuracy through a more precise evaluation of the overall particle
weights $\{W_{l,j}\}$. From a conceptual viewpoint, this choice is motivated
by the fact that, as already mentioned at the beginning of Paragraph
\ref{RBSS_derivation}, the particle $\mathbf{x}_{l/(l-1),j}^{(N)}$ and its
associated Gaussian model $\mathcal{N}(\mathbf{x}_{l}^{(L)};\mathbf{\eta
}_{fp,l,j}^{(L)},\mathbf{C}_{fp,l,j}^{(L)})$ should be considered as two parts
of the same hypothesis, so that they should be exploited \emph{jointly}.

\section{Numerical Results\label{num_results}}

In this Section MPF and the smoothing algorithms developed in this manuscript\footnote{Our 
simulations have evidenced that, for the considered SSM, the Algorithm 2 of \cite[p. 359]%
{Lindsten_2016} suffers from ill-conditioning and that, even if its square root implementation is adopted, its computational load and accuracy are very close to that of the ERBSS technique.} are compared in terms of accuracy and computational load for a specific CLG system, characterized by $D_{L}=3$, $D_{N}=1$ and $P=2$. The structure of the considered system has been 
inspired by the example proposed in \cite{Schon_2010} (where it is proposed as
a good example for the application of MPF) and is characterized by: a) the
state models
\begin{equation}
\mathbf{x}_{l+1}^{(L)}=\left(
\begin{array}
[c]{ccc}%
0.8 & 0.2 & 0\\
0 & 0.7 & -0.2\\
0 & 0.2 & 0.7
\end{array}
\right)  \mathbf{x}_{l}^{(L)}+\left(
\begin{array}
[c]{c}%
\cos(x_{l}^{(N)})\\
-\sin(x_{l}^{(N)})\\
0.5\sin(2x_{l}^{(N)})
\end{array}
\right)  +\mathbf{w}_{l}^{(L)}\label{state_mod_1}%
\end{equation}
and%
\begin{equation}
x_{l+1}^{(N)}=\arctan\left(  x_{l}^{(N)}\right)  +\left(  0.9\quad
0\quad0\right)  \mathbf{x}_{l}^{(L)}+w_{l}^{(N)}\label{state_mod_2}%
\end{equation}
with $\mathbf{w}_{l}^{(L)}\sim\mathcal{N}(0,(\sigma_{w}^{(L)})^{2}%
\mathbf{I}_{3})$, $w_{l}^{(N)}\sim\mathcal{N}(0,(\sigma_{w}^{(N)})^{2}$; b)
the measurement model%
\begin{equation}
\mathbf{y}_{l}=\left(
\begin{array}
[c]{c}%
0.1\left(  x_{l}^{(N)}\right)  {}^{2}\cdot\text{sgn}\left(  x_{l}^{(N)}\right)
\\
0
\end{array}
\right)  +\left(
\begin{array}
[c]{ccc}%
0 & 0 & 0\\
1 & -1 & 1
\end{array}
\right)  \mathbf{x}_{l}^{(L)}+\mathbf{e}_{l}\label{meas_syst}%
\end{equation}
with $\mathbf{e}_{l}\sim\mathcal{N}(0,(\sigma_{e})^{2}\mathbf{I}_{2})$. Note
that the state equation (\ref{state_mod_1}), unlike its counterpart proposed
in \cite{Schon_2010}, depends on $x_{l}^{(N)}$, so that the pseudo-measurement
$\mathbf{z}_{l}^{(N)}$ (\ref{eq:z_N_l}) can be evaluated for this system.

In our computer simulations our assessment of state estimation \emph{accuracy} is based on the evaluation of two \emph{root mean square errors} (RMSEs), one (denoted
$RMSE_{N}($alg$)$, where `alg' denotes the algorithm this parameter refers to)
referring to the (monodimensional) nonlinear state component, the other one
(denoted $RMSE_{L}($alg$)$) to the (three-dimensional) linear state component; note, however, that
the last parameter represents the square root of the average
\emph{mean square error} (MSE) evaluated for the three elements of
$\mathbf{x}_{l}^{(L)}$. Our assessment of \emph{computational
requirements} is based, instead, on assessing the
average \emph{computation time} for processing a single \emph{block} of
measurements (this quantity is denoted CTB in the
following). Moreover, in
our computer simulations, the following choices have been always made: a) $T=200$ has
been selected for the length of the observation interval; b)
$M=N_{p}$ has been chosen for the EBRSS ($M\lesssim N_{p}$ is recommended in
\cite{Lindsten_2013}).

Some results illustrating a) the dependence of $RMSE_{L}$ and $RMSE_{N}$ (CTB) on
the number of particles ($N_{p}$) for the MPF, the RBSS and ERBSS algorithms are illustrated in Fig.
\ref{Fig_5} (Fig. \ref{Fig_6}) \footnote{In these and in the following
figures simulation results are identified by markers, whereas continuous lines
are drawn to ease reading.}; in this case $\sigma_{w}^{(L)}=\sigma_{w}%
^{(N)}=2\cdot10^{-1}$ and $\sigma_{e}=3\cdot10^{-2}$ have been selected. From these
results the following conclusions can be easily inferred for the considered scenario:

\begin{enumerate}
\item On the one hand, a negligible improvement in the estimation accuracy of
all the considered algorithms is achieved for $N_{p}\geq 100$ (actually, a similar result has been found for other values of
$\sigma_{e}$, $\sigma_{w}^{(L)}$ and $\sigma_{w}^{(N)}$); for this reason,
$N_{p}$ $=100$ has been selected in all the computer simulations the following
results refer to.

\item The RBSS algorithm outperforms MPF by about $21.12\%$ ($36.5\%$) in
terms of $RMSE_{L}$ ($RMSE_{N}$) for $N_{p}=100$. A negligible improvement in RBSS accuracy can be obtained by accomplishing a further smoothing for the linear state component (as explained at the end of Paragraph \ref{RBSS_derivation}); this reason, this possibility is no more considered in the following. Note also that the RBSS improvement is obtained at the price of a limited computational cost, since its CTB is about twice that of MPF.

\item The ERBBS algorithm provides a by far richer statistical information
than the RBSS algorithm, but achieves slightly better accuracy in state
estimation and entails a substantially larger computational load, even for small values of
$N_{p}$ (for instance, the ERBBS computation time is about 100 times larger
than that of RBBS for $N_{p}$ $=100$). Note also that the CTB gap between the
EBRSS algorithm and both the RBSS and the MPF techniques becomes larger as
$N_{p}$ increases. For this reason, the ERBBS is not taken into consideration anymore in the
following simulations.

\item A relevant gap between $RMSE_{L}($MPF$)$ and $RMSE_{N}($MPF$)$
($RMSE_{L}($RBSS$)$ and $RMSE_{N}($RBSS$)$) exists; unluckily, the RBSS
algorithm is unable to reduce this gap. This can be
related to the fact that smoothing accuracy is significantly influenced by
that achieved in the forward pass.
\end{enumerate}

A comparison between the MPF and the RBSS state estimation errors has also evidenced that the RMSE improvement provided by the latter algorithm is mainly related to its `peak shaving' effect. In fact, the amplitude of the spikes appearing in the state estimation error at the end of the forward pass are substantially reduced by smoothing. Note, however, that the elements of the system state do not necessarily benefit from this effect in the same way; for instance, for our specific SSM, this effect is stronger for the nonlinear state component than for each of the three elements of the linear state component.

\begin{figure}[ptb]
\vspace{-10mm}
\begin{center}
\includegraphics[scale = 0.50]{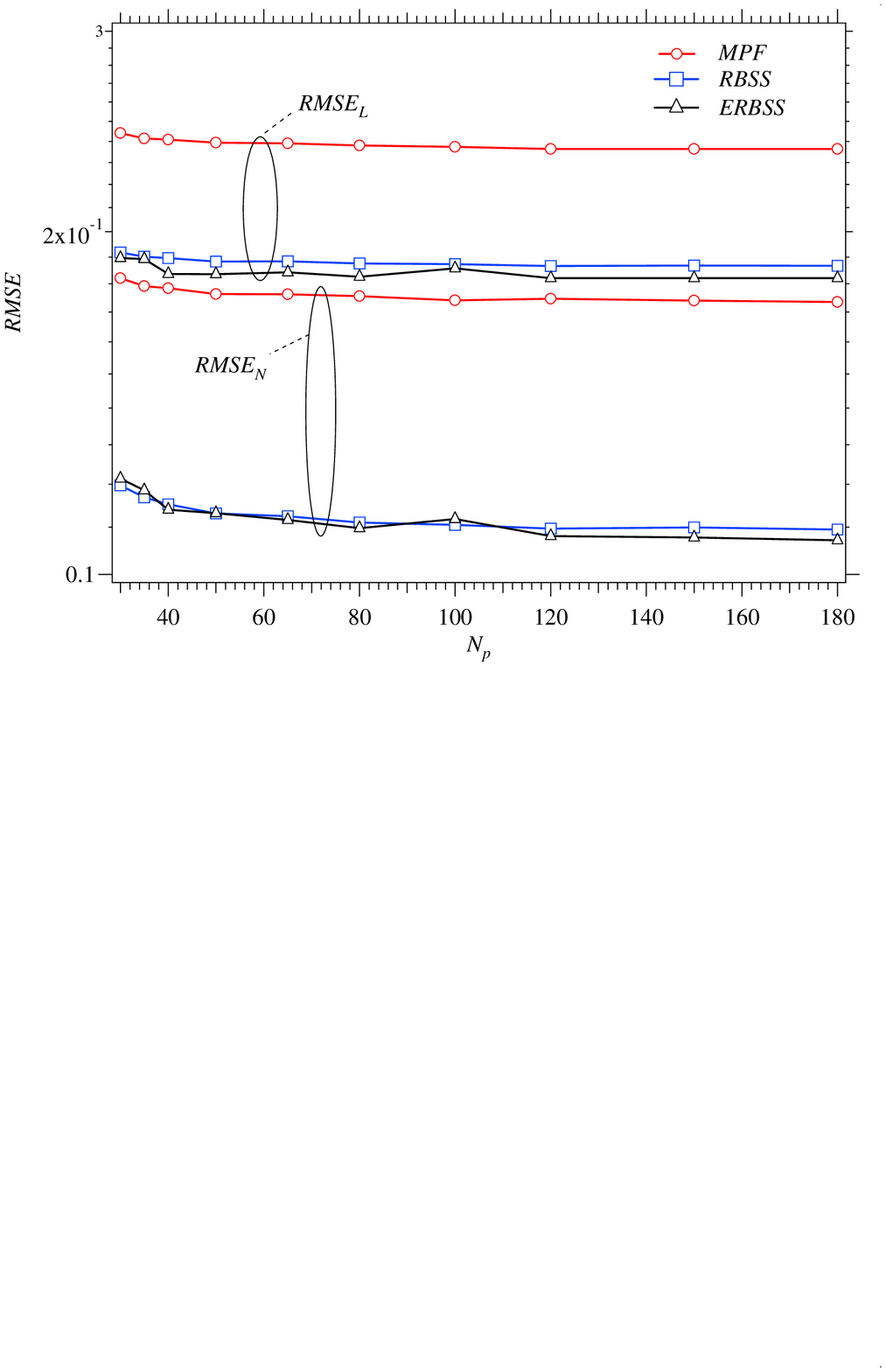}
\end{center}
\vspace{-65mm}
\caption{RMSE performance versus $N_{p}$ for the linear state component
($RMSE_{L}$) and the nonlinear state component ($RMSE_{N}$) for the system
described by eqs. (\ref{state_mod_1})-(\ref{meas_syst}). MPF, RBSS and EBRSS
are considered; $\sigma_{w}^{(L)}=\sigma_{w}^{(N)}=2\cdot10^{-1}$ and
$\sigma_{e}=3\cdot10^{-2}$ have been selected.}%
\label{Fig_5}%
\end{figure}

\begin{figure}[ptb]
\vspace{-10mm}
\begin{center}
\includegraphics[scale = 0.5]{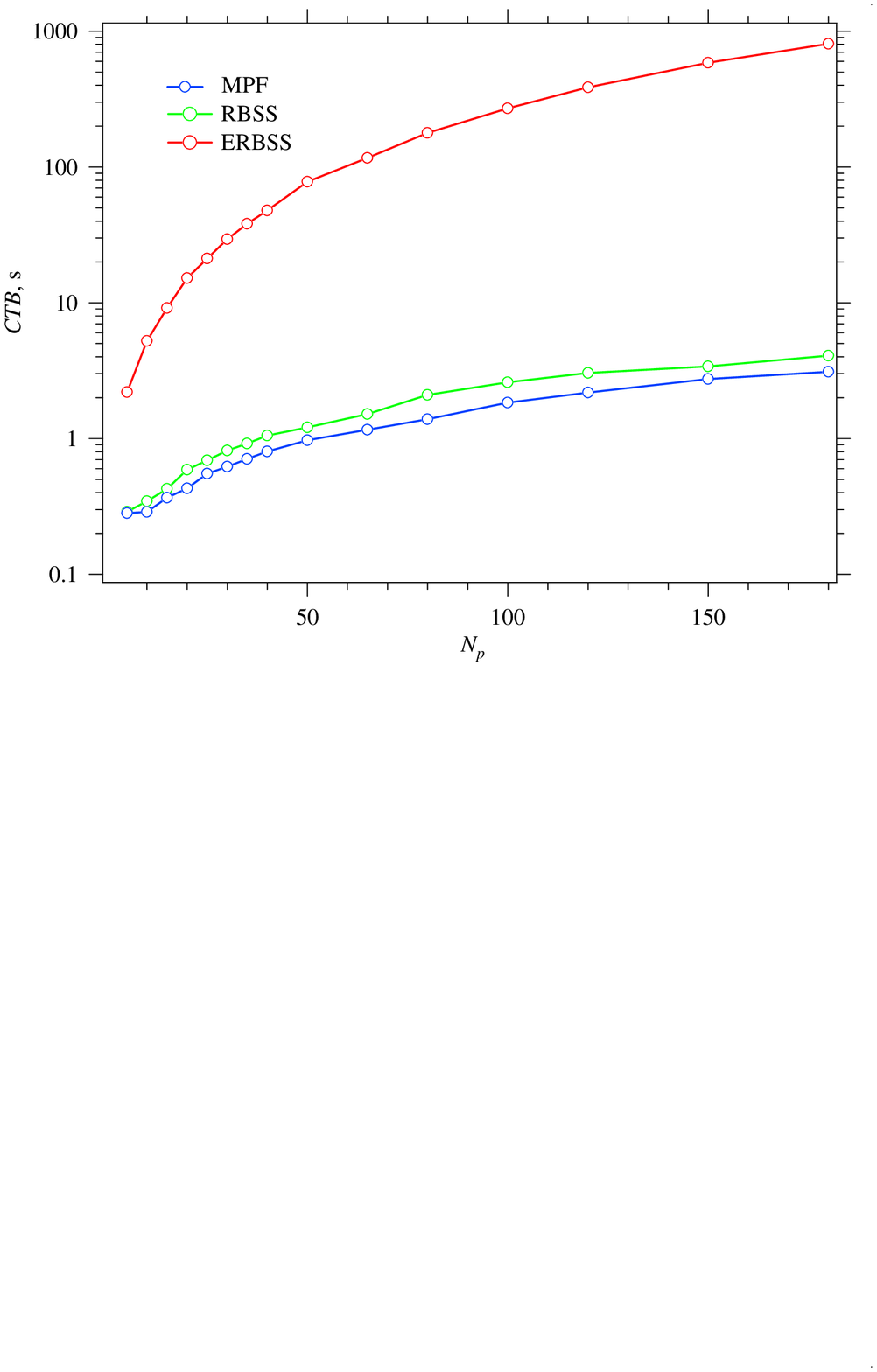}
\end{center}
\vspace{-60mm}
\caption{CTB versus $N_{p}$ for the linear state component ($RMSE_{L}$) and
the nonlinear state component ($RMSE_{N}$) for the system described by eqs.
(\ref{state_mod_1})-(\ref{meas_syst}). MPF, RBSS and EBRSS are considered;
$\sigma_{w}^{(L)}=\sigma_{w}^{(N)}=2\cdot10^{-1}$ and $\sigma_{e}%
=3\cdot10^{-2}$ have been selected.}%
\label{Fig_6}%
\end{figure}

\begin{figure}[ptb]
\vspace{-10mm}
\begin{center}
\includegraphics[scale = 0.50]{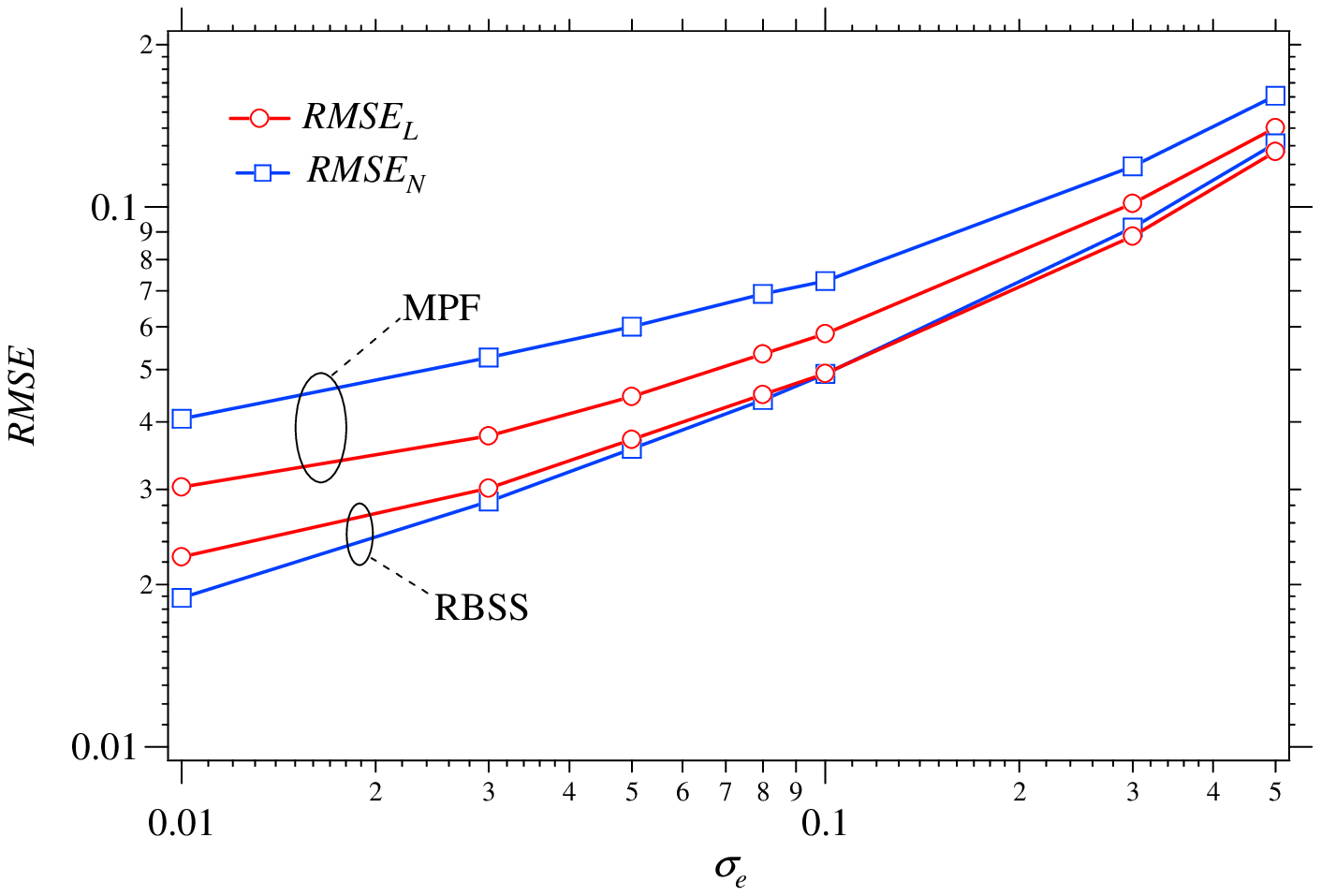}
\end{center}
\vspace{-70mm}
\caption{RMSE performance versus $\sigma_{e}$ for the linear state component
($RMSE_{L}$) and the nonlinear state component ($RMSE_{N}$) for the system
described by eqs. (\ref{state_mod_1})-(\ref{meas_syst}). MPF and RBSS are
considered; $\sigma_{w}^{(L)}=\sigma_{w}^{(N)}=2\cdot10^{-2}$ have been
selected.}
\label{Fig_8}%
\end{figure}

In our work the dependence of $RMSE_{L}$ and $RMSE_{N}$ on the intensity of
the process noise and on that of the measurement noise has been also analysed.
Some results illustrating the dependence of $RMSE_{L}$ and $RMSE_{N}$ on
$\sigma_{e}$ (under the assumption that $\sigma_{w}^{(L)}=$ $\sigma_{w}%
^{(N)}=2\cdot10^{-2}$)\  are shown in Fig. \ref{Fig_8}. From these results it is 
 easily inferred that the performance gap between MPF and RBSS shrinks as
$\sigma_{e}$ increases; this is due to the fact that
 a stronger measurement noise results in a poorer quality of the
statistical information generated in the forward pass, and this impairs more
and more the RBSS estimation process. Other simulation results (not shown here 
for space limitations) have also evidenced that, for a given intensity
of the measurement noise, the gap between $RMSE_{L}($MPF$)$ and $RMSE_{L}%
($RBSS$)$ (and, similarly, between $RMSE_{N}($MPF$)$ and $RMSE_{N}($RBSS$)$)
remains stable as $\sigma_{w}=\sigma_{w}^{(L)}=\sigma_{w}^{(N)}$ changes (in particular, $\sigma_{w}=\in[10^{-2},2\cdot10^{-1}]$ has been assumed in our simulations).


\section{Conclusions\label{sec:conc}}

In this manuscript the smoothing problem for SSMs has been analysed from a FG
perspective. This has allowed us to devise new RBPS methods for CLG\ SSMs. Computer simulations for a specific SSM evidence that
the RBSS algorithm achieves a good performance-complexity tradeoff. Our future
work concerns the application of FG methods to the problems of filtering and
smoothing for other classes of SSMs.

\section*{Acknowledgment}

We would like to thank Dr. Fredrik Lindsten (Uppsala University, Department of Information Technology) for his constructive comments.

{

\end{document}
\begin{thebibliography}{99}                                                                                    
\bibitem {Anderson_1979} B. Anderson and J. Moore, \textbf{Optimal Filtering},
Englewood Cliffs, NJ, Prentice-Hall, 1979.

\bibitem {Arulampalam_2002} M. S. Arulampalam, S. Maskell, N. Gordon and T.
Clapp, \textquotedblleft A Tutorial on Particle Filters for Online
Nonlinear/Non-Gaussian Bayesian Tracking\textquotedblright, \emph{IEEE Trans.
Sig. Proc.}, vol. 50, no. 2, pp. 174-188, Feb. 2002.

\bibitem {Doucet_2001} A. Doucet, J. F. G. de Freitas and N. J. Gordon, ``An
Introduction to Sequential Monte Carlo methods,'' in \textbf{Sequential Monte
Carlo Methods in Practice}, A. Doucet, J. F. G. de Freitas, and N. J. Gordon,
Eds. New York: Springer-Verlag, 2001.

\bibitem {Doucet_2000} A. Doucet, S. Godsill and C. Andrieu, \textquotedblleft
On Sequential Monte Carlo Sampling Methods for Bayesian
Filtering\textquotedblright, \emph{Statist. Comput.}, vol. 10, no. 3, pp.
197-208, 2000.

\bibitem {Gustafsson_2010} F. Gustafsson, \textquotedblleft Particle Filter
Theory and Practice with Positioning Applications\textquotedblright,
\emph{IEEE Aerosp. and Electr. Syst. Mag.}, vol. 25, no. 7, pp. 53-82, July 2010.

\bibitem {Douc_2011} R. Doucet, A. Garivier, E. Moulines and J. Olsson,
\textquotedblleft Sequential Monte Carlo smoothing for general state space
hidden Markov models\textquotedblright, \emph{Ann. Appl. Probab.}, vol. 21,
no. 6, pp. 2109--2145, 2011.

\bibitem {Kitagawa_1987} G. Kitagawa, \textquotedblleft Non-Gaussian
state-space modeling of nonstationary time series\textquotedblright,
\emph{Journal of the American Statistical Association}, vol. 82, pp.
1032-1063, 1987.

\bibitem {Kitagawa_1994} G. Kitagawa, \textquotedblleft The two-filter formula
for smoothing and an implementation of the Gaussian-sum
smoother\textquotedblright, \emph{Annals of the Institute of Statistical
Mathematics}, vol. 46, pp. 605-623, 1994.

\bibitem {Kitagawa_1996} G. Kitagawa, \textquotedblleft Monte Carlo filter and
smoother for non-Gaussian nonlinear state space models\textquotedblright%
,\ \emph{J. Comput. Graph. Statist.}, vol. 5, no. 1, pp. 1--25, 1996.

\bibitem {Bresler_1986} Y. Bresler, \textquotedblleft Two-filter formula for
discrete-time non-linear Bayesian smoothing\textquotedblright%
,\ \emph{Int. Journal of Control}, vol. 43, no. 2, pp. 629-641, 1986.

\bibitem {Vo_2012} B. N. Vo, B. T. Vo and R. P. S. Mahler, \textquotedblleft
Closed-Form Solutions to Forward--Backward Smoothing\textquotedblright,
\emph{IEEE Trans. Sig. Proc.}, vol. 60, no. 1, pp. 2-17, Jan. 2012.

\bibitem {Godsill_2004} S. J. Godsill, A. Doucet, and M. West,
\textquotedblleft Monte Carlo smoothing for nonlinear time
series\textquotedblright, \emph{J. Amer. Statist. Assoc.}, vol. 99, no. 465,
pp. 156--168, Mar. 2004.

\bibitem {Briers_2010} M. Briers, A. Doucet and S. Maskell, \textquotedblleft
Smoothing algorithms for state-space models\textquotedblright, \emph{Ann.
Inst. Statist. Math.}, vol. 62, no. 1, pp. 61--89, Feb. 2010.

\bibitem {Fong_2002} W. Fong, S. J. Godsill, A. Doucet and M. West,
\textquotedblleft Monte Carlo smoothing with application to audio signal
enhancement \textquotedblright, \emph{IEEE Trans. Signal Process.}, vol. 50, no. 2,
pp. 438--449, Feb. 2002.

\bibitem {Lindsten_2013} F. Lindsten and T. B. Sch\"{o}n, \textquotedblleft
Backward simulation methods for Monte Carlo statistical
inference\textquotedblright, \emph{Foundat. Trends Mach. Learn.}, vol. 6, no.
1, pp. 1--143, 2013.

\bibitem {Chen_2000} R. Chen and J. S. Liu, \textquotedblleft Mixture Kalman
filters\textquotedblright, \emph{J. Roy. Statist. Soc.: Ser. B}, vol. 62, no.
3, pp. 493--508, 2000.

\bibitem {Schon_2005} T. Sch\"{o}n, F. Gustafsson, P.-J. Nordlund,
\textquotedblleft Marginalized Particle Filters for Mixed Linear/Nonlinear
State-Space Models\textquotedblright, \emph{IEEE Trans. Sig. Proc.}, vol. 53,
no. 7, pp. 2279-2289, July 2005.

\bibitem {Olsson_2008} J. Olsson, R. Douc, O. Capp\'{e}, and E. Moulines,
\textquotedblleft Sequential Monte Carlo smoothing with application to
parameter estimation in nonlinear state-space models\textquotedblright%
, \emph{Bernoulli}, vol. 14, no. 1, pp. 155--179, 2008.

\bibitem {Lindsten_2016} F. Lindsten, P. Bunch, S. S\"{a}rkk\"{a}, T. B.
Sch\"{o}n and S. J. Godsill, \textquotedblleft Rao-Blackwellized Particle
Smoothers for Conditionally Linear Gaussian Models\textquotedblright,
\emph{IEEE J. Sel. Topics in Sig. Proc.}, vol. 10, no. 2, pp. 353-365, March 2016.

\bibitem {Vitetta_2016} G. M. Vitetta, E. Sirignano, F. Montorsi and M. Sola,
\textquotedblleft Marginalized Particle Filtering and Related Filtering
Techniques as Message Passing\textquotedblright, submitted to the
\emph{IEEE\ Trans. Inf. Theory}, july 2016 (available online at https://arxiv.org/abs/1605.03017).

\bibitem {Vitetta_2017} G. M. Vitetta, E. Sirignano and F. Montorsi,
\textquotedblleft A Novel Message Passing Algorithm for Online
Bayesian Filtering: Turbo Filtering\textquotedblright, to be presented at the
\emph{IEEE ICC 2017 Workshop on Advances in Network Localization and Navigation (ANLN)}, May 2017.

\bibitem {Loeliger_2007} H.-A. Loeliger, J. Dauwels, Junli Hu, S. Korl, Li
Ping, F. R. Kschischang, ``The Factor Graph Approach to Model-Based Signal
Processing'', \emph{IEEE Proc.}, vol. 95, no. 6, pp. 1295-1322, June 2007.

\bibitem {Kschischang_2001} F. R. Kschischang, B. Frey, and H. Loeliger,
\textquotedblleft Factor Graphs and the Sum-Product
Algorithm\textquotedblright, \emph{IEEE Trans. Inf. Theory}, vol. 41, no. 2,
pp. 498-519, Feb. 2001.

\bibitem {Loeliger_2016} H.-A. Loeliger, L. Bruderer, H. Malmberg, F. Wadehn
and N. Zalmai, \textquotedblleft On Sparsity by NUV-EM, Gaussian Message
Passing, and Kalman Smoothing\textquotedblright, \emph{Proc. of the}
\emph{2016 Inf. Theory \& Appl. Workshop} (ITA), La Jolla, CA
(USA), Jan. 2016.

\bibitem {Wadehn_2016} F. Wadehn, J. Dauwels, H.-A. Loeliger and H. Yu,
\textquotedblleft Outlier-insensitive Kalman Smoothing and Marginal Message
Passing\textquotedblright, \emph{Proc. of the 24th European Sig. Proc.
Conf.} (EUSIPCO 2016), Budapest (Hungary), August 2016.

\bibitem {Schon_2010} T. Sch\"{o}n, \textquotedblleft Example Used in
Exemplifying the Marginalized (Rao-Blackwellized) Particle Filter", Nov. 2010
(available at http://users.isy.liu.se/en/rt/schon/Code/RBPF/Document/MPFexample.pdf).

\bibitem {Runnalls_2007} A. R. Runnalls, \textquotedblleft Kullback-Leibler
Approach to Gaussian Mixture Reduction\textquotedblright, \emph{IEEE Trans. on
Aerosp. and Elec. Syst.}, vol. 43, no. 3, pp. 989-999, July 2007.

\end{thebibliography}
